\documentclass[twocolumn,amsmath]{autart}    % Enable this line and disable the 
\usepackage{graphicx}          % Include this line if your 
\usepackage{bm}                             % bold italic vector symbols
\usepackage{acronym}
\usepackage{amssymb,amsfonts}
\usepackage{mathtools}
\usepackage[inline]{enumitem}
\usepackage[per-mode=symbol]{siunitx} % Consistent physical units. `per-mode=fraction` for fraction mode (e.g., m/s instead of ms^-1). For S column type that aligns numbers by the decimal point and allows specifying column widths.
\usepackage{subfig}
\usepackage{booktabs}   % \toprule, \midrule, \bottomrule
\usepackage{makecell}   % \makecell for multi-line table cells
\usepackage{url}

% ...
% ...
% ...
% ...
% ...
% ...
\begin{document}

\begin{frontmatter}

\title{TTCBF: A Truncated Taylor Control Barrier Function for High-Order Safety Constraints\thanksref{footnoteinfo}} % Title, preferably not more 

\thanks[footnoteinfo]{This paper was not presented at any IFAC 
meeting. Corresponding author J.~Xu. Tel. +49 241 80-21149. 
Fax +49 241 80-22150.}

\author[Paestum]{Jianye Xu}\ead{xu@embedded.rwth-aachen.de},    % Add the 
\author[Rome]{Bassam Alrifaee}\ead{bassam.alrifaee@unibw.de},               % e-mail address 
% \author[Baiae]{Publius Maro Vergilius}\ead{vergilius@culture.ir}  % (ead) as shown

\address[Paestum]{Chair of Embedded Software, RWTH Aachen University, Aachen 52072, Germany}  % Please supply                                              
\address[Rome]{Department of Aerospace Engineering, University of the Bundeswehr Munich, Munich 85579, Germany}             % full addresses
% \address[Baiae]{The White House, Baiae}        % here.

                             % keyword list or with the 
                                          % help of the Automatica 
                                          % keyword wizard

% ===============================
% Acronym definitions
% ===============================
% Case-insensitive acronym helpers
\acrodef{ACC}{Adaptive Cruise Control}

\acrodef{BF}{Barrier Function}

\acrodef{CAV}{Connected and Automated Vehicle}
\acrodef{CBF}{Control Barrier Function}
\acrodef{CLF}{Control Lyapunov Function}
\acrodef{CPM}{Cyber-Physical Mobility}
\acrodef{CPMLab}{Cyber-Physical Mobility Lab}

\acrodef{DCBF}{Discrete-Time Control Barrier Function}

\acrodef{eCBF}{exponential \acs{CBF}}
\acrodef{eHOCBF}{exponential \acs{HOCBF}}
\acrodef{ESP}{Elementary Symmetric Polynomial}

\acrodef{HOCBF}{High-Order \acs{CBF}}

\acrodef{LP}{Linear Program}

\acrodef{MPC}{Model Predictive Control}

\acrodef{NLP}{Non-Linear Program}

\acrodef{OCP}{Optimal Control Problem}
\acrodef{ODE}{Ordinary Differential Equation}

\acrodef{PACBF}{Parameter-Adaptive \acs{CBF}}

\acrodef{QP}{Quadratic Program}

\acrodef{RACBF}{Relaxation-Adaptive \acs{CBF}}

\acrodef{TTCBF}{Truncated Taylor \acs{CBF}}
\acrodef{aTTCBF}{adaptive \acs{TTCBF}}

\begin{abstract} % Abstract of not more than 200 words.
Control Barrier Functions (CBFs) enforce safety by rendering a prescribed safe set forward invariant. However, standard CBFs are limited to safety constraints with relative degree one, while High-Order CBF (HOCBF) methods address higher relative degree at the cost of introducing a chain of auxiliary functions and multiple class $\mathcal{K}$ functions whose tuning scales with the relative degree. In this paper, we introduce a Truncated Taylor Control Barrier Function (TTCBF), which generalizes standard discrete-time CBFs to consider high-order safety constraints and requires only one class $\mathcal{K}$ function, independent of the relative degree. 
We also propose an adaptive variant, adaptive TTCBF (aTTCBF), that optimizes an online gain on the class $\mathcal{K}$ function to improve adaptability, while requiring fewer control design parameters than existing adaptive HOCBF variants. Numerical experiments in a relative-degree-six spring-mass system and a cluttered corridor navigation validate the above theoretical findings.

\end{abstract}

\end{frontmatter}

\acresetall
%===============================================================================
\section{Introduction}\label{sec:introduction}
\acp{CBF} are a standard tool for safety-critical control, where safety is specified as forward invariance of a set defined by a barrier inequality \cite{ames2014control}. In the \ac{CBF} framework, one encodes a state constraint through a continuously differentiable function and derives a pointwise inequality constraint on the control input whose satisfaction implies forward invariance of the safe set. When combined with a \ac{CLF} constraint and a quadratic objective, this construction yields a sequence of \acp{QP} that can be solved online, giving rise to the widely used \ac{CLF}-\ac{CBF}-\ac{QP} framework \cite{ames2017control,du2023reinforcement}. This framework has been demonstrated in applications such as autonomous driving \cite{ames2017control,ames2019control,xu2025realtime} and safe navigation for mobile robots \cite{dai2023safe,xu2025learningbased}, where it provides real-time implementability together with closed-loop safety guarantees.

\begin{figure}[t]
    \centering
    \includegraphics[width=1.0\linewidth]{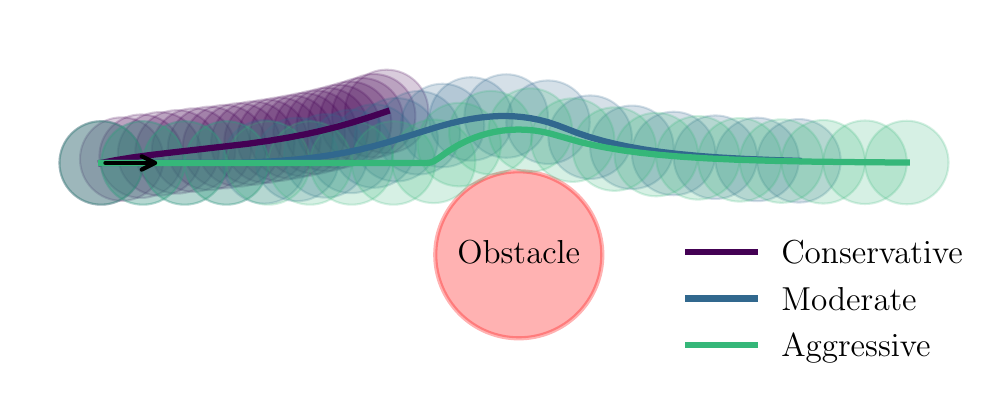}
    \caption{An obstacle-avoidance example with three different parameters of a class $\mathcal{K}$ function: conservative, moderate, and aggressive. Footprints: circles; trajectories: solid lines.}
    \label{fig_example_class_k}
\end{figure}

Class $\mathcal{K}$ functions play a central role in \acp{CBF} because they regulate how the barrier value evolves as the state approaches the safety boundary. Their parameters directly affect the trade-off between conservatism and aggressiveness: conservative choices can shrink the feasible set of the associated \ac{QP} and lead to infeasibility under tight control bounds \cite{jankovic2018robust}, whereas aggressive choices can allow rapid motion toward the boundary and increase the likelihood of infeasibility when the state reaches the boundary. Fig.~\ref{fig_example_class_k} illustrates how this tuning changes the resulting obstacle-avoidance behavior, ranging from slow and conservative to fast and aggressive maneuvers. 

A further limitation of standard \acp{CBF} is their reliance on relative-degree-one constraints. The relative degree of a safety constraint with respect to a system is the number of times one needs to differentiate it along the system dynamics before the control input appears explicitly. For higher-relative-degree constraints, early approaches include backstepping-based constructions \cite{hsu2015control} and exponential \acp{CBF} based on input--output linearization \cite{nguyen2016exponential}. The \ac{HOCBF} in \cite{xiao2022highorder} provides a constructive extension that handles arbitrarily high relative degree by introducing a chain of auxiliary functions. This construction uses one class $\mathcal{K}$ function at each level of the chain, so the number of class $\mathcal{K}$ functions, and hence the number of associated tuning parameters, grows with the relative degree. Several methods aim to reduce this tuning burden or improve feasibility, including penalty and parameterization methods \cite{xiao2022highorder}, adaptive variants such as \ac{PACBF} and \ac{RACBF} \cite{xiao2022adaptive}, sampled-data adaptive \acp{CBF} \cite{xiong2023discretetime}, and learning-based designs of class $\mathcal{K}$ functions \cite{ma2022learning,kim2025learning}. Among sampled-data safety formulations, the zero-order \ac{CBF} in \cite{tan2025zeroorder} compares barrier values at consecutive sampling instants and predicts the next-step value through numerical approximation of the system flow. This enables high-relative-degree constraints but makes the safety condition implicit and sensitive to the accuracy of flow prediction.

We address the control-design complexity induced by high-relative-degree constraints and the resulting increase in tuning parameters for class $\mathcal{K}$ functions. 
Our main contributions are:
\begin{enumerate}[leftmargin=*]
    \item We propose \ac{TTCBF}, a new \ac{CBF} variant that accommodates high-relative-degree safety constraints using only one class $\mathcal{K}$ function, independent of the relative degree. We show that our \ac{TTCBF} is a generalization of the standard discrete-time \ac{CBF}, which can only consider safety constraints with relative degree one.
    \item We further propose \ac{aTTCBF}, an adaptive variant of our \ac{TTCBF} that improves adaptability while requiring substantially fewer control-design parameters than existing adaptive variants of the standard \ac{HOCBF}.
\end{enumerate}
We validate our \ac{TTCBF} and \ac{aTTCBF} through numerical experiments with a spring-mass system with relative degree six and a corridor-navigation scenario with densely cluttered obstacles.

We adopt the following notation throughout the paper. Vectors are written in boldface. The indices $i\in\mathbb{N}$ and $j\in\mathbb{N}$ are reused with meanings defined locally. The index $k\in\mathbb{N}$ denotes discrete time steps, while $t\coloneqq k\Delta t\in\mathbb{R}$ denotes a time instant with $\Delta t>0$ being the sampling period of the controller.

The remainder of the paper is organized as follows.
Section~\ref{sec:preliminaries} introduces preliminaries for \acp{HOCBF}.
Section~\ref{sec:taylor-based-formulation} introduces our \ac{TTCBF} and our \ac{aTTCBF} based on truncated Taylor expansions, which handle high-relative-degree safety constraints using only one class $\mathcal{K}$ function.
Section~\ref{sec:experiments} reports numerical results, followed by discussions in Section~\ref{sec:discussions} and conclusions in Section~\ref{sec:conclusions}.

%===============================================================================
%===============================================================================
\section{Preliminaries} \label{sec:preliminaries}
%==========================
We consider an input-affine control system of the form
\begin{equation} \label{eq:affine-sys}
    \dot{\bm{x}} = f(\bm{x}) + g(\bm{x})\bm{u},
\end{equation}
where the system state is $\bm{x} \coloneqq [x_1,\ldots,x_n]^\top \in \mathcal{X} \subset \mathbb{R}^n$.
The vector fields $f:\mathbb{R}^n \to \mathbb{R}^n$ and $g:\mathbb{R}^n \to \mathbb{R}^{n \times m}$ are assumed to be locally Lipschitz.
The control input $\bm{u} \coloneqq [u_1,\ldots,u_m]^\top \in \mathcal{U} \subseteq \mathbb{R}^m$ is componentwise constrained as
$\mathcal{U} \coloneqq \bigl\{ \bm{u} \in \mathbb{R}^m \mid \bm{u}_{\min} \le \bm{u} \le \bm{u}_{\max} \bigr\}$,
where $\bm{u}_{\min}, \bm{u}_{\max} \in \mathbb{R}^m$ denote the minimum and maximum admissible control inputs, respectively.

In safety-critical control, the objective is to ensure that the system state $\bm{x}$ remains within a prescribed (possibly time-varying) safe set $\mathcal{C}(t)$.
Let $h:\mathcal{X} \times [t_0,\infty) \to \mathbb{R}$ be a continuously differentiable safety function, where $t_0\in\mathbb{R}$ denotes an initial time.
The safe set is defined as
\[
\mathcal{C}(t) \coloneqq \bigl\{ \bm{x} \in \mathcal{X} \mid h(\bm{x},t) \ge 0 \bigr\}.
\]

For system \eqref{eq:affine-sys}, define the gradient of $h$ with respect to $\bm{x}$ as
$
\nabla_{\bm{x}} h(\bm{x},t) \coloneqq
\begin{bmatrix}
\frac{\partial h}{\partial x_1}, \ldots, \frac{\partial h}{\partial x_n}
\end{bmatrix}^{\top}.
$
The Lie derivatives of $h$ along $f$ and $g$ are given by
\begin{align}
L_f h(\bm{x},t) &\coloneqq \nabla_{\bm{x}} h(\bm{x},t)^{\top} f(\bm{x}), \\
L_g h(\bm{x},t) &\coloneqq \nabla_{\bm{x}} h(\bm{x},t)^{\top} g(\bm{x}).
\end{align}

For a time-varying safety function $h(\bm{x},t)$, its time derivative along trajectories of \eqref{eq:affine-sys} is
\[
\dot{h}(\bm{x},\bm{u},t)
= \frac{\partial h(\bm{x},t)}{\partial t}
+ L_f h(\bm{x},t)
+ L_g h(\bm{x},t)\bm{u}.
\]
Higher-order time derivatives $\frac{d^i}{dt^i}h(\bm{x}(t),t)$ are defined recursively by differentiating along trajectories of \eqref{eq:affine-sys}. Since standard \ac{CBF} formulations require the control input to appear in the first derivative of the safety function, they are not applicable to safety constraints with higher relative degree. To formalize this notion, we adopt the following definition.
\begin{defn}[Relative Degree {\cite{xiao2022highorder}}] \label{def:relative-degree}
The relative degree $r\in\mathbb{N}$ of a sufficiently differentiable function
$h:\mathcal{X}\times[t_0,\infty)\to\mathbb{R}$
with respect to system \eqref{eq:affine-sys} is the smallest integer such that
\[
\frac{\partial}{\partial \bm{u}}
\left(\frac{d^i}{dt^i}h(\bm{x}(t),t)\right)= \bm{0}_{1\times m},
\quad \forall i\in\{0,\ldots,r-1\}.
\]
\[
\frac{\partial}{\partial \bm{u}}
\left(\frac{d^r}{dt^r}h(\bm{x}(t),t)\right)\neq \bm{0}_{1 \times m},
\]
\end{defn}
\begin{defn}[Class $\mathcal{K}$ function~\cite{xiao2022highorder}]
    A function $\alpha:[0,b)\to[0,\infty)$, with $b>0$, belongs to class $\mathcal{K}$ if it is Lipschitz continuous, strictly increasing, and $\alpha(0)=0$.
\end{defn}
For a safety function $h$ with relative degree $r$, the \ac{HOCBF} proposed in \cite{xiao2022highorder} introduces a sequence of auxiliary functions
\begin{equation} \label{eq:chain-hocbf}
\begin{aligned}
    &\Psi_0(\bm{x},t) \coloneqq h(\bm{x},t), \\
    &\begin{aligned}
        \Psi_i(\bm{x},t) \coloneqq
        \dot{\Psi}_{i-1}(\bm{x},t)
        + &\alpha_i\left(\Psi_{i-1}(\bm{x},t)\right), \\
        &\qquad \quad \forall i \in \{1,\ldots,r-1\}, 
    \end{aligned}\\
    &\Psi_r(\bm{x},\bm{u},t) \coloneqq
    \dot{\Psi}_{r-1}(\bm{x},\bm{u},t)
    + \alpha_r\!\left(\Psi_{r-1}(\bm{x},t)\right),
\end{aligned}
\end{equation}
where each $\alpha_i(\cdot)$ is a class $\mathcal{K}$ function.
Here, $\dot{\Psi}_i$ denotes the time derivative of $\Psi_i$ along the system dynamics \eqref{eq:affine-sys}.
By construction, the control input $\bm{u}$ appears explicitly only in $\Psi_r$.

Based on \eqref{eq:chain-hocbf}, define the sets
\begin{equation} \label{eq:high-order-sets}
\mathcal{C}_i(t) \coloneqq \bigl\{ \bm{x} \in \mathcal{X} \mid \Psi_{i-1}(\bm{x},t) \ge 0 \bigr\}, \forall i \in \{1,\ldots,r\}.
\end{equation}
% $\partial \mathcal{C}_i(t) \coloneqq \bigl\{ \bm{x} \in \mathcal{X} \mid \Psi_{i-1}(\bm{x},t) = 0$
A set is said to be \textit{forward invariant} for system \eqref{eq:affine-sys}
if every trajectory starting in the set remains in the set for all future time \cite{xiao2022highorder}.

\begin{defn}[\acp{HOCBF} {\cite{xiao2022highorder}}] \label{def:continuous-t-hocbf}
A sufficiently differentiable function
$h:\mathcal{X} \times [t_0,\infty) \to \mathbb{R}$
is an \ac{HOCBF} of relative degree $r$ for system \eqref{eq:affine-sys}
if there exist class $\mathcal{K}$ functions $\alpha_i(\cdot)$, $i \in \{1,\ldots,r\}$, such that
\begin{equation} \label{eq:hocbf-cond-simple}
    \sup_{\bm{u} \in \mathcal{U}} \Psi_r(\bm{x},\bm{u},t) \ge 0,
    \quad
    \forall (\bm{x},t) \in
    \bigcap_{i=1}^{r} \mathcal{C}_i(t) \times [t_0,\infty).
\end{equation}
\end{defn}

Using \eqref{eq:chain-hocbf}, $\Psi_r$ admits the explicit expression
\begin{equation}
\begin{aligned}
\Psi_r(\bm{x},\bm{u},t)
= &\mathcal{L}^r h(\bm{x},t) + L_g \mathcal{L}^{r-1} h(\bm{x},t)\bm{u} \\
&+ \sum_{i=1}^{r} \mathcal{L}^{i-1}
\alpha_{r-i+1}\!\left(\Psi_{r-i}(\bm{x},t)\right),
\end{aligned}
\end{equation}
where $\mathcal{L} \coloneqq \partial_t + L_f$ denotes a differential operator, with $\partial_t$ being the partial derivative with respect to time.

Since $\Psi_0 \coloneqq h$, the set $\mathcal{C}_1$ coincides with the original safe set.
As shown in \cite{xiao2022highorder}, enforcing condition \eqref{eq:hocbf-cond-simple}
ensures that $\Psi_i(\bm{x},t) \ge 0$ for all $i \in \{0,\ldots,r-1\}$,
which renders the set $\bigcap_{i=1}^{r} \mathcal{C}_i(t)$ forward invariant
for system \eqref{eq:affine-sys}.

%====================================================================
\section{Truncated Taylor-Based Formulation} \label{sec:taylor-based-formulation}

We propose our \ac{TTCBF} in Section~\ref{sec:ttcbf} and \ac{aTTCBF} in Section~\ref{sec:attcbf}. 

\subsection{Truncated Taylor CBF (TTCBF)} \label{sec:ttcbf}
The motivation for our \ac{TTCBF} is that the standard \ac{HOCBF} method proposed in \cite{xiao2022highorder} requires $r$ class $\mathcal{K}$ functions, where $r$ is the relative degree of the safety constraint. The simplest choice is a linear class $\mathcal{K}$ function $\alpha(h)=ah$ with $a>0$ as a tuning parameter. Even with this choice, \ac{HOCBF} still requires $r$ class-$\mathcal{K}$ parameters. This directly couples the number of tuning parameters with the relative degree of the safety constraint. Since the parameters of class $\mathcal{K}$ functions directly determine the tightness of the safety constraint and the aggressiveness of the resulting control behavior (as demonstrated in Fig.~\ref{fig_example_class_k}), improper tuning can lead to degraded control performance or excessive conservatism. To simplify control synthesis and parameter tuning, we introduce our \ac{TTCBF}, which decouples the number of tuning parameters from the relative degree.

Before we introduce our \ac{TTCBF}, we define the notion of the relative degree in the discrete-time domain. Let $k \in \mathbb{N}$ denote the time step index and $t_k \coloneqq k\Delta t$ the time instant, where $\Delta t > 0$ is the sampling period of the digital controller.
\begin{defn}[\cite{monaco1987minimumphase}]\label{def:relative-degree-discrete-time}
Let $y(t_k)=h\bigl(\bm{x}(t_k),t_k\bigr)$ be a system output at time step $k$ with
$h:\mathcal{X}\times[t_0,\infty)\to\mathbb{R}$.
The relative degree $r\in\mathbb{N}$ of $y$ is the smallest integer such that
\begin{equation}
\begin{aligned}
\frac{\partial y(t_{k+i})}{\partial \bm{u}(t_k)}
&= \bm{0}_{1\times m}, \quad \forall i\in\{0,\ldots,r-1\},\\
\frac{\partial y(t_{k+r})}{\partial \bm{u}(t_k)}
&\neq \bm{0}_{1\times m}.
\end{aligned}
\end{equation}
\end{defn}

Informally, the relative degree $r$ of a system output in discrete time is the number of time steps after which the control input $\bm{u}(t_k)$ first affects the output $y$.
For systems with relative degree $r=1$, the standard discrete-time \ac{CBF} condition takes the form
\begin{equation} \label{eq:standard-discrete-t-cbf}
h\bigl(\bm{x}(t_{k+1}), t_{k+1}\bigr)
- h\bigl(\bm{x}(t_k), t_k\bigr)
+ \alpha\bigl(h\bigl(\bm{x}(t_k), t_k\bigr)\bigr)
\ge 0,
\end{equation}
where $\alpha(\cdot)$ is a class~$\mathcal{K}$ function satisfying $\alpha(s) \le s$ for all $s \ge 0$ \cite{agrawal2017discrete}.
For outputs with relative degree $r>1$, \cite{xiong2023discretetime} extends this condition by cascading $r$ class~$\mathcal{K}$ functions, in direct analogy to the continuous-time \ac{HOCBF} construction in \cite{xiao2022highorder}.
Below, we present a direct $r$-step generalization that involves only one class~$\mathcal{K}$ function.
Before proceeding, we introduce the following proposition.

\begin{prop}\label{prop:r-step-cbf}
Denote the discrete-time safe set
\begin{equation}\label{eq:safe-set-discrete-t}
\mathcal{C}(t_k)\coloneqq\{\bm x(t_k)\in\mathcal X\mid h(\bm x(t_k),t_k)\ge0\}
\end{equation}
for system~\eqref{eq:affine-sys}. Let $k_0\in\mathbb N$ be an initial time step, and assume the initial conditions
\begin{equation}\label{eq:initial-cond-safe}
h\bigl(\bm x(t_{k_0+i}),t_{k_0+i}\bigr)\ge0,\quad \forall i\in\{0,\ldots,r-1\}.
\end{equation}
Let $\alpha(\cdot)$ be a class $\mathcal K$ function that satisfies $\alpha(s)\le s$ for all $s\ge0$.
Assume that for every $k\ge k_0$ with $\bm x(t_k)\in\mathcal{C}(t_k)$, the controller applies an input
$\bm u(t_k)\in\mathcal U$ such that
\begin{equation}\label{eq:r-step-cbf}
h\bigl(\bm x(t_{k+r}),t_{k+r}\bigr)-h\bigl(\bm x(t_k),t_k\bigr)
+\alpha\Bigl(h\bigl(\bm x(t_k),t_k\bigr)\Bigr)\ge0,
\end{equation}
where $\bm x(t_{k+r})$ is the closed-loop state at time $t_{k+r}$ under the applied inputs.
Then $\mathcal{C}(t_k)$ is forward invariant for system \eqref{eq:affine-sys}, that is,
$h\bigl(\bm x(t_k),t_k\bigr)\ge0$ for all $k\ge k_0$.
\end{prop}

\begin{pf}
We prove $h\bigl(\bm x(t_k),t_k\bigr)\ge0$ for all $k\ge k_0$ by induction.
The base case for $k\in\{k_0,\ldots,k_0+r-1\}$ holds by~\eqref{eq:initial-cond-safe}.
For the inductive step, fix any $j\ge k_0+r-1$ and assume
$h\bigl(\bm x(t_k),t_k\bigr)\ge0$ for all $k\le j$.
Let $k'\coloneqq j+1-r$, so that $k'\ge k_0$ and $k'+r=j+1$.
Applying~\eqref{eq:r-step-cbf} at time step $k'$ yields
$h\bigl(\bm x(t_{j+1}),t_{j+1}\bigr)
=h\bigl(\bm x(t_{k'+r}),t_{k'+r}\bigr)
\ge h\bigl(\bm x(t_{k'}),t_{k'}\bigr)
-\alpha\Bigl(h\bigl(\bm x(t_{k'}),t_{k'}\bigr)\Bigr).$
By the induction hypothesis, $h\bigl(\bm x(t_{k'}),t_{k'}\bigr)\ge0$.
Using $\alpha(s)\le s$ for $s\ge0$ gives
$h\bigl(\bm x(t_{k'}),t_{k'}\bigr)-\alpha(h(\bm x(t_{k'}),t_{k'}))\ge0$,
and therefore $h\bigl(\bm x(t_{j+1}),t_{j+1}\bigr)\ge0$.
This completes the induction.
\qed
\end{pf}

\begin{rem}\label{rem:initial-r-steps}
For $r>1$, Definition~\ref{def:relative-degree-discrete-time} implies that the control input chosen at time $t_{k_0}$
does not affect $y(t_{k_0+i})=h(\bm x(t_{k_0+i}),t_{k_0+i})$ for $i\in\{1,\ldots,r-1\}$.
Hence, condition~\eqref{eq:initial-cond-safe} cannot be enforced by the control applied at time $t_{k_0}$.
This role of initial nonnegativity is analogous to requiring nonnegative initial values of the auxiliary variables
$\Psi_i(\bm x,t)$, $i\in\{0,\ldots,r-1\}$, in the \ac{HOCBF} construction~\eqref{eq:chain-hocbf}.
\end{rem}

$h\bigl(\bm x(t_{k+r}),\bm u(t_k),t_{k+r}\bigr)$ in \eqref{eq:r-step-cbf} is generally not an explicit function of the current control input~$\bm u(t_k)$. A common approximation is to apply successive forward differences to obtain
$h\bigl(\bm x(t_{k+i}),t_{k+i}\bigr)$ from $i=1$ until $i=r$ \cite{xiong2023discretetime}. However, for non-linear systems, this often yields a non-affine dependence on~$\bm u(t_k)$ that requires further approximation. Additionally,  \cite{xiong2023discretetime} relies on a cascade of class~$\mathcal K$ functions that require tuning. We circumvent these problems by expanding $h\bigl(\bm x(t_{k+r}),\bm u(t_k),t_{k+r}\bigr)$ in a \emph{truncated Taylor series} about $h\bigl(\bm x(t_k),t_k\bigr)$. Let the sampling period of the controller be $\Delta t>0$ and define the Taylor step size 
\begin{equation} \label{eq:taylor-size}
    \Delta T \coloneqq r\Delta t>0.
\end{equation}
Taylor's theorem gives
\begin{equation}\label{eq:taylor-expansion}
\begin{aligned}
  & h\bigl(\bm x(t_{k+r}),\bm u(t_k),t_{k+r}\bigr) \\
   = &h\bigl(\bm x(t_k),t_k\bigr) + \sum_{i=1}^{r-1}\frac{(\Delta T)^i}{i!} h^{(i)}\bigl(\bm x(t_k),t_k\bigr) + \\
  &\frac{(\Delta T)^r}{r!} h^{(r)}\bigl(\bm x(t_k),\bm u(t_k),t_k\bigr) + R_{\mathrm T},
\end{aligned}
\end{equation}
where the \emph{remainder} $R_\mathrm{T}$ admits the Lagrange form
\begin{equation} \label{eq:taylor-remainder}
  R_{\mathrm T} \!=\!
  \frac{(\Delta T)^{r+1}}{(r+1)!}
  h^{(r+1)}\bigl(\bm x(\tau),
                   \bm u(\tau),
                   \tau\bigr),
\end{equation}
with $\tau\!\in\!(t_k,t_{k+r})$ being an intermediate time. We expand the Taylor series until the $r$-th derivative term because this is where $\bm u$ explicitly appears. We choose $\Delta T$ as in \eqref{eq:taylor-size} because the influence of~$\bm u(t_k)$ to $h$ is delayed by $r$ time steps according to Definition~\ref{def:relative-degree-discrete-time}. Now, we define our \ac{TTCBF} as follows. 
\begin{defn}
[\acl{TTCBF} (\acs{TTCBF})]\label{def:ttcbf}
Let $h:\mathcal X\times[t_0,\infty)\to\mathbb R$ be an $(r+1)$-times continuously differentiable function that defines a safe set $\mathcal{C}(t_k)$ as in \eqref{eq:safe-set-discrete-t} for system~\eqref{eq:affine-sys}. Let $k_0 \in \mathbb{N}$ be an initial time step, and let the initial conditions \eqref{eq:initial-cond-safe} hold. Then, $h$ is called a \ac{TTCBF} of relative degree~$r$ for system~\eqref{eq:affine-sys} if there exists a class~$\mathcal K$  function $\alpha(\cdot)$ such that
  \begin{equation}\label{eq:ttcbf-cond}
  \begin{aligned}
    \sup_{\bm u(t_k)\in\mathcal U}
    \Biggl[
      &\sum_{i=1}^{r-1}\frac{(\Delta T)^i}{i!}
        h^{(i)}\bigl(\bm x(t_k),t_k\bigr) + \\
        &\quad \frac{(\Delta T)^r}{r!}h^{(r)}\bigl(\bm x(t_k),\bm u(t_k),t_k\bigr) + \\
        & \quad \qquad R_{\mathrm T}
      + \alpha\bigl(h\bigl(\bm x(t_k),t_k\bigr)\bigr)
    \Biggr] \ge 0
  \end{aligned}
  \end{equation}
holds for every $k \ge k_0$ with $\bm x(t_k)\in \mathcal{C}(t_k)$.
\end{defn}

In \eqref{eq:ttcbf-cond}, the remainder $R_\mathrm{T}$ is unknown since the intermediate time~$\tau$ in \eqref{eq:taylor-remainder} is unknown. We tackle this problem with the following theorem.
\begin{thm}\label{thm:ttcbf}
  Let $h$ be a \ac{TTCBF} with relative degree $r$ for system \eqref{eq:affine-sys} with the associated safe set~$\mathcal{C}(t_k)$ as in~Definition~\ref{def:ttcbf}. Let $\Delta T$ be a Taylor step size as in \eqref{eq:taylor-size}. Any Lipschitz continuous controller satisfying
  \begin{align} \label{eq:thm-ttcbf-cond}
    \sum_{i=1}^{r-1}\frac{(\Delta T)^i}{i!}&h^{(i)}\bigl(\bm x(t_k),t_k\bigr)
    + \frac{(\Delta T)^r}{r!}h^{(r)}\bigl(\bm x(t_k),\bm u(t_k),t_k\bigr) \notag \\
    &+ \alpha\bigl(h\bigl(\bm x(t_k),t_k\bigr)\bigr) + \inf_{\bm u(t_k)\in\mathcal U}R_{\mathrm T}
    \ge 0
  \end{align}
  renders~$\mathcal{C}(t_k)$ forward invariant for
  system~\eqref{eq:affine-sys}.
\end{thm}

\begin{pf}
Substituting \eqref{eq:taylor-expansion} into \eqref{eq:r-step-cbf} yields 
    \begin{align} \label{eq:ttcbf-cond-approx}
    \sum_{i=1}^{r-1}\frac{(\Delta T)^i}{i!}
      & h^{(i)}\bigl(\bm x(t_k),t_k\bigr) + \frac{(\Delta T)^r}{r!}h^{(r)}\bigl(\bm x(t_k),\bm u(t_k),t_k\bigr) \notag \\
    &+ \alpha\bigl(h\bigl(\bm x(t_k),t_k\bigr)\bigr) + R_{\mathrm T}
    \ge 0.
    \end{align}
    Since \eqref{eq:thm-ttcbf-cond} considers the worst-case (most negative) $R_{\mathrm T}$ over all admissible control inputs $\mathcal{U}$, thus \eqref{eq:thm-ttcbf-cond} is a sufficient condition for \eqref{eq:ttcbf-cond-approx}. Given Proposition~\ref{prop:r-step-cbf}, $\mathcal{C}(t_k)$ is forward invariant for system \eqref{eq:affine-sys}.
\qed
\end{pf}

To compute the term $\inf_{\bm u(t_k)\in\mathcal U}R_{\mathrm T}$ in \eqref{eq:thm-ttcbf-cond}, we denote
$
  h^{(i)}_{k,\min}\coloneqq
  \inf_{\bm u(t_k)\in\mathcal U}
  h^{(i)}\bigl(\bm x(t_k),\bm u(t_k),t_k\bigr)
$
for $i\in\{r,r+1\}$.
Given \eqref{eq:taylor-remainder}, applying the backward difference $h^{(r+1)}_{k,\min}=(h^{(r)}_{k,\min}-h^{(r)}_{k-1}) / \Delta t$ yields
\begin{equation}\label{eq:approx-taylor-remainder}
\begin{aligned}
  \inf_{\bm u(t_k)\in\mathcal U}R_{\mathrm T}
  &= \frac{(\Delta T)^{r+1}}{(r+1)!} h^{(r+1)}_{k,\min} \\
  &= \frac{(\Delta T)^{r+1}}{(r+1)!\Delta t} \Bigl(h^{(r)}_{k,\min}-h^{(r)}_{k-1}\Bigr),
\end{aligned}
\end{equation}
where $h^{(r)}_{k-1}$ denotes the $r$-th time derivative of $h$ at the previous time step $k-1$, which is known at the current time~$k$. Next, we discuss how to determine $h_{k,\min}^{(r)}$. For input-affine systems like \eqref{eq:affine-sys}, the $r$-th time derivative of $h$ depends on the control input $\bm u(t_k)$ in an affine manner. Consequently, computing $h_{k,\min}^{(r)}$ reduces to minimizing an affine function of $\bm u(t_k)$ over the admissible set $\mathcal{U}$. If $\mathcal{U}$ is a polytope (defined by linear inequalities), this minimization is a linear program and can be solved efficiently online. If $\mathcal{U}$ is a norm ball (e.g., $\|\bm{u}\|_2\le u_{\max}$), minimizing an affine function over $\mathcal{U}$ often admits a closed-form solution \cite[Appendix A.1.6]{boyd2004convex}. However, if the dependence on $\bm u$ is non-affine or if $\mathcal{U}$ is defined by nonlinear nonconvex constraints, the minimization becomes a generally nonconvex nonlinear program, which limits the applicability of our approach in such cases.

\begin{rem}
For safety constraints with relative degree $r = 1$, \eqref{eq:initial-cond-safe} becomes $h\bigl(\bm x(t_{k_0}),t_{k_0}\bigr) \ge 0$ and \eqref{eq:r-step-cbf} reduces to the standard discrete-time \ac{CBF} condition \eqref{eq:standard-discrete-t-cbf} proposed in \cite{agrawal2017discrete}. Therefore, our \ac{TTCBF} generalizes the standard discrete-time \ac{CBF} to high-order safety constraints.
\end{rem}

% The consideration of the worst-case Taylor remainder $R_{\mathrm{T}}$ introduces conservatism. However, in our numerical experiments, this conservatism was unnoticeable, owing to its exponentially scaling factor $(\Delta T)^{r+1}/(r+1)!$.

%====================================================================
\subsection{Adaptive TTCBF (aTTCBF)}\label{sec:attcbf}
Safety-critical controllers often operate in changing environments. Adaptive \acp{CBF} address this setting by adjusting the class $\mathcal{K}$ function online, which in turn modifies the closed-loop behavior. For \ac{HOCBF}, however, the number of class $\mathcal{K}$ functions scales with the relative degree of the safety constraint, so the control-design complexity of existing adaptive variants of \ac{HOCBF} also scales with the relative degree. In contrast, our \ac{TTCBF} uses only one class $\mathcal{K}$ function, which simplifies the construction of adaptive variants. In this section, we introduce an adaptive variant of our \ac{TTCBF}, termed \ac{aTTCBF}.
\begin{defn}
[\acl{aTTCBF} (\acs{aTTCBF})]\label{def:attcbf}
Let $h:\mathcal X\times[t_0,\infty)\to\mathbb R$ be an $(r+1)$-times continuously differentiable function that defines a safe set $\mathcal{C}(t_k)$ as in \eqref{eq:safe-set-discrete-t} for system~\eqref{eq:affine-sys}. Let $k_0 \in \mathbb{N}$ be an initial time step, and let the initial condition \eqref{eq:initial-cond-safe} hold. Then, $h$ is called an \ac{aTTCBF} of relative degree~$r$ for system~\eqref{eq:affine-sys} if there exists a class~$\mathcal K$  function $\hat \alpha(\cdot)$ such that
  \begin{equation}\label{eq:attcbf-cond}
  \begin{aligned}
    \sup_{\bm u(t_k)\in\mathcal U}
    \Biggl[
      &\sum_{i=1}^{r-1}\frac{(\Delta T)^i}{i!}
        h^{(i)}\bigl(\bm x(t_k),t_k\bigr) + \\
        &\quad \frac{(\Delta T)^r}{r!}h^{(r)}\bigl(\bm x(t_k),\bm u(t_k),t_k\bigr) + \\
        & \quad \qquad R_{\mathrm T}
      + \underbrace{\eta(t_k)\hat{\alpha}\bigl(h(\bm{x}(t_k),t_k)\bigr)}_{\alpha(h,t_k)}
    \Biggr] \ge 0
  \end{aligned}
  \end{equation}
  holds for every $k \ge k_0$ with $(\bm x(t_k),t_k)\in \mathcal{C}(t_k) \times [t_{k_0},\infty)$. Here, $\hat{\alpha}(\cdot)$ is a coefficient-free class~$\mathcal{K}$ function satisfying $\hat{\alpha}(h) \le h$, i.e., whose scaling gain is normalized to one, and $\eta(t_k)\in [0,1]$ is an adaptive gain.
\end{defn}

Essentially, we separate the class $\mathcal{K}$ function $\alpha(h,t_k)$ into two multiplicative parts: a gain $\eta(t_k)$ and a coefficient-free class~$\mathcal{K}$ function $\hat{\alpha}(\cdot)$. Conventionally, this gain is a parameter that must be manually tuned. In our \ac{aTTCBF}, we use an adaptive gain $\eta(t_k)$ that is optimized online. We have a theorem for our \ac{aTTCBF} similar to Theorem~\ref{thm:ttcbf} for our \ac{TTCBF} as follows.

\begin{thm}\label{thm:attcbf}
  Let $h$ be an \ac{aTTCBF} with relative degree $r$ for system \eqref{eq:affine-sys} with the associated safe set~$\mathcal{C}(t_k)$ as in~Definition~\ref{def:attcbf}. Let $\Delta T$ be a Taylor step size as in \eqref{eq:taylor-size}. Any Lipschitz continuous controller satisfying
  \begin{align} \label{eq:thm-attcbf-cond}
    \sum_{i=1}^{r-1}&\frac{(\Delta T)^i}{i!}h^{(i)}\bigl(\bm x(t_k),t_k\bigr)
    + \frac{(\Delta T)^r}{r!}h^{(r)}\bigl(\bm x(t_k),\bm u(t_k),t_k\bigr) \notag \\
    &+ \eta(t_k)\hat\alpha\bigl(h\bigl(\bm x(t_k),t_k\bigr)\bigr) + \inf_{\bm u(t_k)\in\mathcal U}R_{\mathrm T}
    \ge 0
  \end{align}
  renders~$\mathcal{C}(t_k)$ forward invariant for
  system~\eqref{eq:affine-sys}.
\end{thm}
\begin{pf}
Since $\hat{\alpha}(\cdot)$ satisfies $\hat{\alpha}(h) \le h$ and $\eta(t_k) \in [0,1]$, the inequality $\alpha(h,t_k)\le h$ holds, which is required by the standard discrete-time \acp{CBF} \eqref{eq:standard-discrete-t-cbf}. The remaining proof follows directly from the proof of Theorem~\ref{thm:ttcbf}.
\qed
\end{pf}
% For example, when using a linear class $\mathcal{K}$ function, we can use $\alpha(h,t_k)=\eta(t_k)\,h$ with $\eta(t_k) \in [0,1]$. More examples (exponential and rational class $\mathcal{K}$ functions) can be found in Section~\ref{sec:eva-corridor}. 

\subsection{Optimal Control Problem Formulation}\label{sec:ocp}
We assume that a nominal controller exists whose safety we seek to certify, while minimizing the deviation from its behavior. For our \ac{TTCBF} and \ac{aTTCBF}, we formulate a \ac{CLF}-\ac{CBF}-\ac{QP} \cite{ames2017control} that is solved at each time step $k$. For clarity, we present the case with one safety constraint. One can extend the formulation to multiple safety constraints by adding the corresponding safety constraints to the \ac{QP} and augmenting the cost with additional penalty terms.

\begin{subequations}\label{eq:qp}
\begin{align}
&
\begin{aligned}
\mathop{\mathrm{arg\,min}}\limits_{\substack{
    \bm u(t_k)\in\mathcal{U},\\
    \zeta(t_k)\ge 0,\\
    \eta(t_k)\in[0,1]}}\!\!
\|\bm u\!(t_k)\!-\!\bm u_{\mathrm{nom}}\!(t_k)\|_{W_u}^2
      \!+\! w_\zeta \zeta(t_k)^2 \! \underbrace{+ w_\eta \eta(t_k)^2}_{\mathrm{aTTCBF\ only}}
\end{aligned}\label{eq:qp-cost}\\
&\text{s.t.}\notag\\
& \ \text{Safety: \eqref{eq:thm-ttcbf-cond} for \ac{TTCBF}, or \eqref{eq:thm-attcbf-cond} for \ac{aTTCBF}}\label{eq:qp-safety}\\
& \ \text{Stability: } \dot V(\bm x(t_k))+\alpha_{\mathrm{L}}\big(V(\bm x(t_k))\big) \le \zeta(t_k)\label{eq:qp-stability}
\end{align}
\end{subequations}
In \eqref{eq:qp-cost}, $W_u\in\mathbb{R}^{m\times m}\succ 0$ penalizes the deviation of the optimized control $\bm u(t_k)$ from the nominal input $\bm u_{\mathrm{nom}}(t_k)$. If a nominal controller is unavailable, one can omit $\bm u_{\mathrm{nom}}(t_k)$ so that this term penalizes control effort. The slack variable $\zeta(t_k)\ge 0$ relaxes the \ac{CLF} constraint \eqref{eq:qp-stability} and is penalized by $w_\zeta>0$. For \ac{aTTCBF}, the additional term with weight $w_\eta>0$ penalizes the adaptive gain $\eta(t_k)\in[0,1]$ in \eqref{eq:thm-attcbf-cond}. Alternatively, one can introduce a nominal adaptive gain and penalize deviations from it. Note that the $\eta(t_k)$ only exists in \eqref{eq:qp-cost} for \ac{aTTCBF}. The stability constraint \eqref{eq:qp-stability} enforces convergence of $\bm x(t_k)$ toward a desired state $\bm x_{\mathrm{des}}(t_k)$ via a \ac{CLF}, where $\alpha_{\mathrm{L}}(\cdot)$ is a class $\mathcal{K}$ function. A common choice is
$V(\bm x(t_k))=\|\bm x(t_k)-\bm x_{\mathrm{des}}(t_k)\|_2^2$.
Without loss of generality, we assume that the \ac{CLF} has relative degree one. If this assumption does not hold, one can use a high-order \ac{CLF} constructed analogously to high-order \acp{CBF}, which we ignore for brevity.

\begin{rem}[Adaptability of \ac{aTTCBF}]\label{rem:adaptability-attcbf}
The adaptability of our \ac{aTTCBF} arises from the decision variable $\eta(t_k)$ in \eqref{eq:qp}. It acts as an adaptive gain of the class $\mathcal{K}$ function in \eqref{eq:thm-attcbf-cond}. The quadratic penalty $w_\eta \eta(t_k)^2$ encourages small values of $\eta(t_k)$ while allowing $\eta(t_k)$ to increase when necessary to maintain feasibility. When \eqref{eq:thm-attcbf-cond} becomes infeasible under the control bounds $\mathcal{U}$, increasing $\eta(t_k)$ relaxes the constraint. Conversely, when the safety constraint is inactive, for example, when the system state is far from the safety boundary, the optimizer reduces $\eta(t_k)$ to limit the allowed rate at which the system state approaches the safety boundary, that is, to limit the decay rate of the barrier value. This mechanism provides a continuous adjustment of conservatism that can improve feasibility under tight control bounds.
\end{rem}

\section{Numerical Experiments}\label{sec:experiments}
We conduct numerical experiments to support our theoretical results and to validate the proposed \ac{TTCBF} and \ac{aTTCBF}. In Section~\ref{sec:eva-spring-mass}, we demonstrate that our \ac{TTCBF} can handle safety constraints with high relative degree using a serial spring-mass system whose safety constraint has relative degree six. In Section~\ref{sec:eva-corridor}, we illustrate the applicability of our \ac{TTCBF} and \ac{aTTCBF} to different types of class $\mathcal{K}$ functions in a corridor-navigation scenario. In the same scenario, we compare our \ac{aTTCBF} with \ac{PACBF} and \ac{RACBF}, two adaptive variants proposed in \cite{xiao2022adaptive} for \ac{HOCBF} \cite{xiao2022highorder}, and show improved adaptability of our \ac{aTTCBF}. All simulations are implemented in Python and executed on an Apple M2 Pro with 16~GB RAM. We formulate and solve the \acp{QP} using \texttt{CVXPY} \cite{diamond2016cvxpy}. The code to reproduce our results, together with video demonstrations, is publicly available\footnote{\url{https://github.com/bassamlab/ttcbf}}.

\subsection{Spring-Mass System: Relative Degree Six}\label{sec:eva-spring-mass}
\begin{figure}[t]
    \centering

    \subfloat[Spring-mass system setup.\label{fig_spring_mass_system}]{
        \begin{minipage}[t]{\linewidth}
            \centering
            \includegraphics[width=\linewidth]{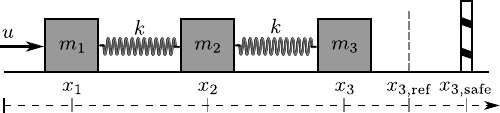}
        \end{minipage}
    }

    \vspace{-3mm}

    \subfloat[Positions of the three masses $m_1$, $m_2$, and $m_3$.\label{fig_spring_mass_position}]{
        \begin{minipage}[t]{\linewidth}
            \centering
            \includegraphics[width=\linewidth]{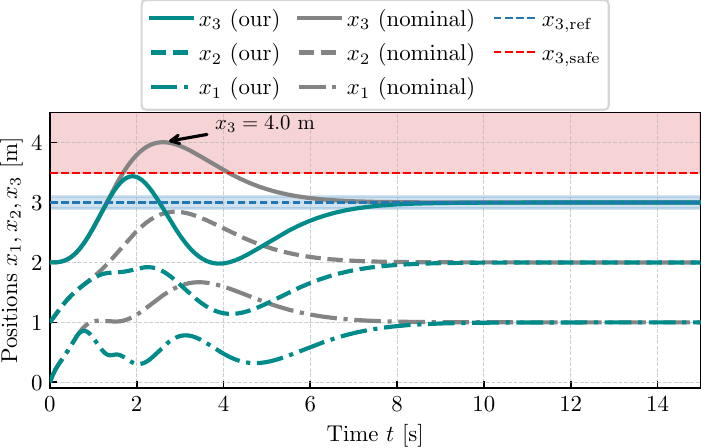}
        \end{minipage}
    }

    \vspace{-3mm}

    \subfloat[Control input $u$ applied by the nominal and our controller.\label{fig_spring_mass_force}]{
        \begin{minipage}[t]{\linewidth}
            \centering
            \includegraphics[width=\linewidth]{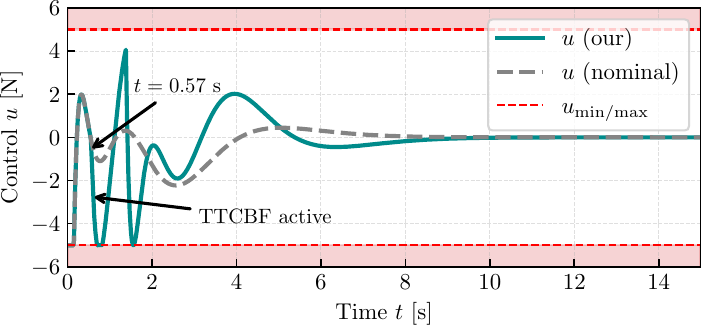}
        \end{minipage}
    }
    \vspace{-2mm}
    \caption{
        Spring-mass system controlled by the nominal and our controller. Forbidden areas are shown in red. 
    }
    \label{fig_spring_mass}
\end{figure}

We evaluate our \ac{TTCBF} on the serial spring-mass benchmark introduced in \cite{nguyen2016exponential}, which features a safety constraint with relative degree six. Fig.~\ref{fig_spring_mass_system} illustrates the setup. Three identical point masses $m_1,m_2, \text{and } m_3$ of \SI{1.0}{\kilogram} move along a horizontal line and are connected by identical linear springs of stiffness $k=\SI{5.0}{\newton\per\meter}$. An external force $u$ bounded by $\lvert u\rvert\le\SI{5}{\newton}$ acts on the first mass $m_1$. Let $x_i$ denote the horizontal position of each mass $i$. The system dynamics are
\begin{equation*}
    \begin{aligned}
        m_1 \ddot{x}_1 &= u + k(x_2 - x_1),\notag\\
        m_2 \ddot{x}_2 &= k(x_1 - x_2) + k(x_3 - x_2),\notag\\
        m_3 \ddot{x}_3 &= k(x_2 - x_3).\notag\\
    \end{aligned}
\end{equation*}
With the state vector $\bm{x}=[x_1,x_2,x_3,\dot{x}_1,\dot{x}_2,\dot{x}_3]^\top \in \mathbb{R}^6$, the system is linear time-invariant, $\dot{\bm{x}}=A\bm{x}+Bu$, where
\begin{equation*}\label{eq:sm_state}
A\!=\!\begin{bmatrix}
\bm 0_{3\times 3} &\!\!\!\! I_3\\
A' &\!\!\!\! \bm 0_{3\times 3}
\end{bmatrix},
A'\!=\!\begin{bmatrix}
-\tfrac{k}{m_1} &\!\!\!\! \tfrac{k}{m_1} &\!\!\!\! 0\\
\tfrac{k}{m_2} &\!\!\!\! -\tfrac{2k}{m_2} &\!\!\!\! \tfrac{k}{m_2}\\
0 &\!\!\!\! \tfrac{k}{m_3} &\!\!\!\! -\tfrac{k}{m_3}
\end{bmatrix},
B\!=\!\begin{bmatrix}
\bm 0_{3\times 1}\\
\tfrac{1}{m_1}\\
\bm 0_{2\times 1}
\end{bmatrix},
\end{equation*}
with $I_3$ denoting an identity matrix of size 3.

The control objective is to steer the third mass to a desired position $x_{3,\mathrm{des}}=\SI{3.0}{\meter}$ while enforcing the safety constraint $x_3\le x_{3,\mathrm{safe}}=\SI{3.5}{\meter}$. We define the safe set as $\mathcal{C}=\{\bm{x}: h(\bm{x})\coloneqq x_{3,\mathrm{safe}}-x_3\ge0\}$. Therefore, the input $u$ must propagate through two springs and three double integrators. One can easily show that $h$ has relative degree six with respect to the system dynamics. We use sampling period $\Delta t=\SI{0.01}{\second}$ and set the initial state $\bm{x}_0=[0,1,2,2,1,0]^\top$.

To assess the effectiveness of our \ac{TTCBF}, we implement two controllers: 1) a \textit{nominal controller} with I/O-linearization that aims at following $x_{3,\mathrm{des}}$ and ignores safety, and 2) a \ac{CBF}-\ac{QP}-based controller employing our \ac{TTCBF} that aims to certify the safety of the nominal controller, referred to as \textit{our controller} thereafter. 
We design the nominal controller using exact input-output linearization followed by pole placement. Let the system output be $y \coloneqq C_y \bm x=x_3$, where $C_y=[0,0,1,0,0,0]$. Since $y$ has relative degree six, the input-output dynamics can be expressed as $y^{(6)} = C_yA^6\bm{x} + C_yA^5Bu$.   Defining the tracking error $e^{(i)} \coloneqq y^{(i)} - y_{\mathrm{ref}}^{(i)}$ with output reference $y_{\mathrm{ref}}=x_{3,\mathrm{des}}$, we specify the desired error dynamics as $e^{(6)} + a_5 e^{(5)} + a_4 e^{(4)} + a_3 e^{(3)} + a_2 e^{(2)} + a_1 e^{(1)} + a_0 e = 0$, where the coefficients $a_i$ correspond to the expanded form of $(s+\lambda)^6$ for a chosen pole location $\lambda>0$ (which we tune as $\lambda=2.0$). The resulting control law is
\begin{equation*}
    u = \frac{e^{(6)} - C_yA^6\bm{x}}{C_yA^5B}, \text{ where } e^{(6)} = -\sum_{i=0}^5 a_i\,e^{(i)}.
\end{equation*}
In our controller, we define $\mathrm{arg\,min}_{u \in \mathcal{U}}(u - u_{\mathrm{nom}})^2$ as the cost function, which optimizes the control $u$ such that it minimally deviates from the nominal control input $u_\mathrm{nom}$ while respecting the safety constraint \eqref{eq:ttcbf-cond-approx} imposed by our \ac{TTCBF}. For \eqref{eq:ttcbf-cond-approx}, we apply a linear class~$\mathcal{K}$ function $\alpha(h)=\lambda h$ with $\lambda=0.95$ (after proper tuning).

\textbf{Results}: Fig.~\ref{fig_spring_mass_position} depicts the positions of each mass $m_i$, and the position of $m_3$ is shown in solid lines. Both controllers are able to track the desired position $x_{3,\mathrm{des}}$ for mass $m_3$. However, the nominal controller (gray) overshoots and reaches $x_3=\SI{4.0}{\meter}$ at $t=\SI{2.6}{\second}$, thereby violating the safety bound by $\SI{0.4}{\meter}$. Our controller (cyan) successfully prevents this violation. Fig.~\ref{fig_spring_mass_force} shows the applied control $u$. Up to $t=\SI{0.57}{\second}$, both controllers apply the same input because the nominal control is safe, and thus our \ac{TTCBF} is inactive. After this time instance, it becomes active and reduces $u$ so that mass $m_1$ pushes $m_2$ less aggressively, and $m_3$ approaches the safety boundary gently, reaching $x_{3,\mathrm{safe}}$ at $t=\SI{1.9}{\second}$ without overshoot.

\begin{figure*}[t]
    \centering
    % ---------- Row 1: Trajectories ----------
    \subfloat[Trajectory (linear class $\mathcal{K}$)\label{fig_corridor_attcbf_linear_trajectories}]{
        \begin{minipage}[b]{0.32\textwidth}
            \centering
            \includegraphics[width=\linewidth]{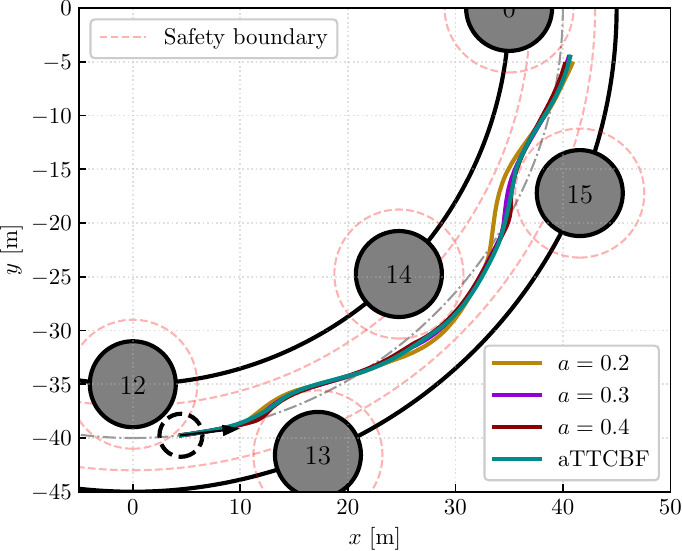}
        \end{minipage}
    }\hfill
    \subfloat[Trajectory (exponential class $\mathcal{K}$)\label{fig_corridor_attcbf_exp_trajectories}]{
        \begin{minipage}[b]{0.32\textwidth}
            \centering
            \includegraphics[width=\linewidth]{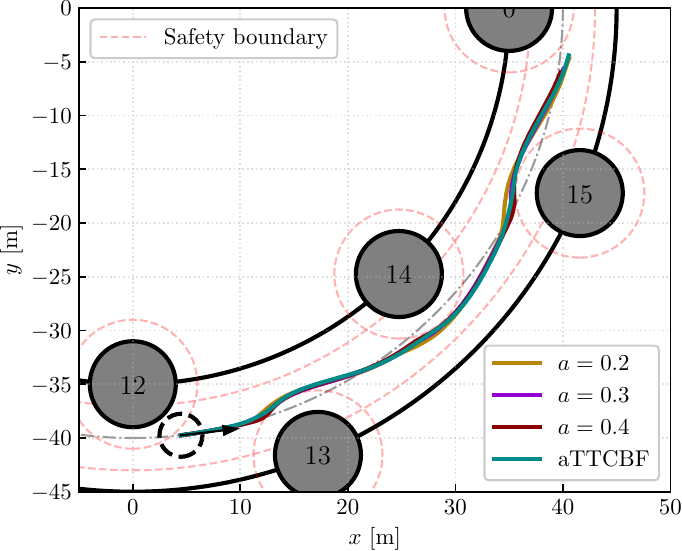}
        \end{minipage}
    }\hfill
    \subfloat[Trajectory (rational class $\mathcal{K}$)\label{fig_corridor_attcbf_rational_trajectories}]{
        \begin{minipage}[b]{0.32\textwidth}
            \centering
            \includegraphics[width=\linewidth]{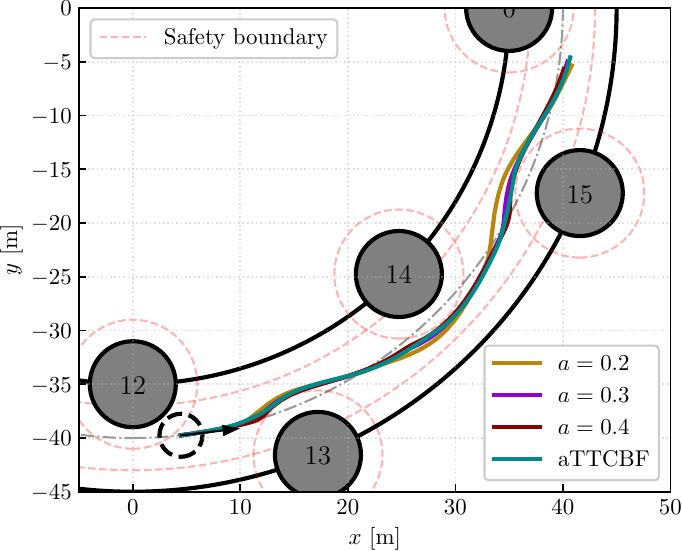}
        \end{minipage}
    }

    \vspace{-3mm}
    
    % ---------- Row 2: u1 Figures ----------
    \subfloat[Steering rate $u_1$ (linear $\mathcal{K}$)\label{fig_corridor_attcbf_linear_u1}]{
        \begin{minipage}[b]{0.32\textwidth}
            \centering
            \includegraphics[width=\linewidth]{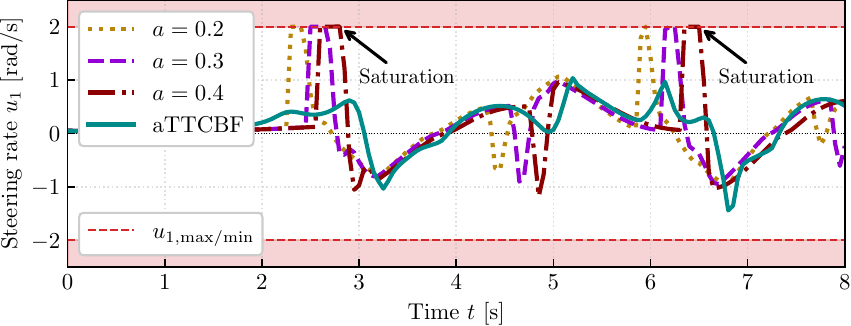}
        \end{minipage}
    }\hfill
    \subfloat[Steering rate $u_1$ (exp. $\mathcal{K}$)\label{fig_corridor_attcbf_exp_u1}]{
        \begin{minipage}[b]{0.32\textwidth}
            \centering
            \includegraphics[width=\linewidth]{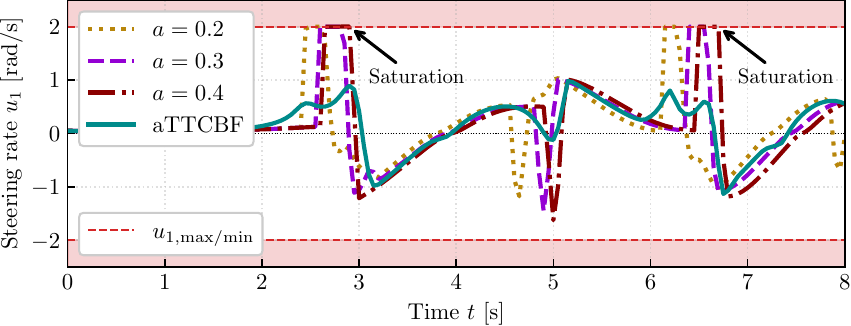}
        \end{minipage}
    }\hfill
    \subfloat[Steering rate $u_1$ (rational $\mathcal{K}$)\label{fig_corridor_attcbf_rational_u1}]{
        \begin{minipage}[b]{0.32\textwidth}
            \centering
            \includegraphics[width=\linewidth]{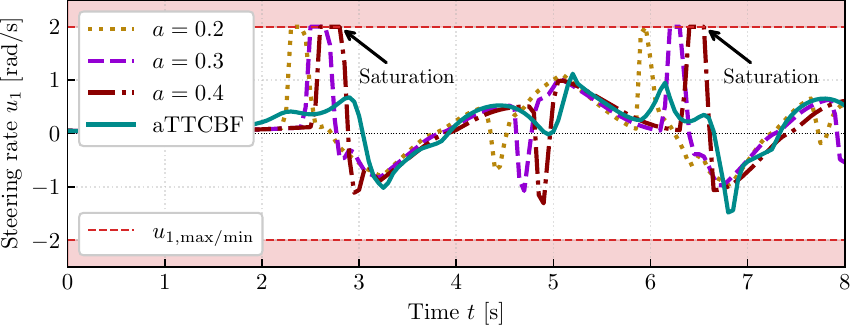}
        \end{minipage}
    }

    \vspace{-3mm}

    % ---------- Row 3: u2 Figures ----------
    \subfloat[Acceleration $u_2$ (linear $\mathcal{K}$)\label{fig_corridor_attcbf_linear_u2}]{
        \begin{minipage}[b]{0.32\textwidth}
            \centering
            \includegraphics[width=\linewidth]{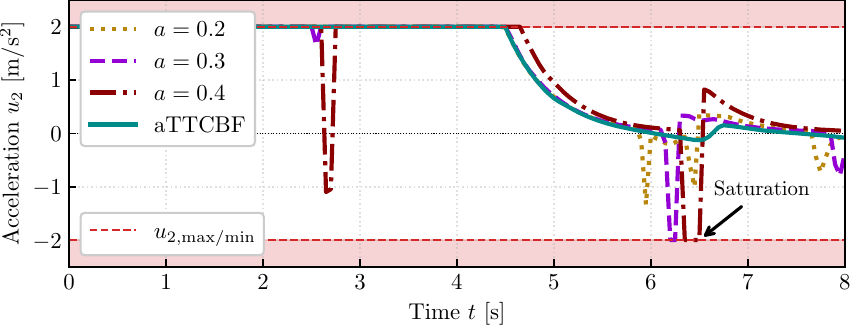}
        \end{minipage}
    }\hfill
    \subfloat[Acceleration $u_2$ (exp. $\mathcal{K}$)\label{fig_corridor_attcbf_exp_u2}]{
        \begin{minipage}[b]{0.32\textwidth}
            \centering
            \includegraphics[width=\linewidth]{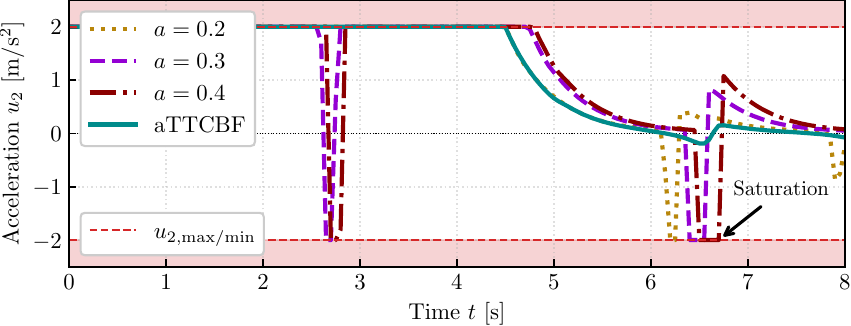}
        \end{minipage}
    }\hfill
    \subfloat[Acceleration $u_2$ (rational $\mathcal{K}$)\label{fig_corridor_attcbf_rational_u2}]{
        \begin{minipage}[b]{0.32\textwidth}
            \centering
            \includegraphics[width=\linewidth]{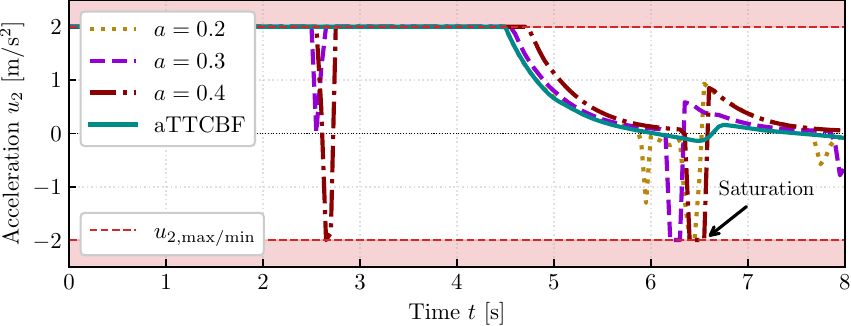}
        \end{minipage}
    }

    \vspace{-3mm}

    % ---------- Row 4: v Figures ----------
    \subfloat[Speed $v$ (linear $\mathcal{K}$)\label{fig_corridor_attcbf_linear_v}]{
        \begin{minipage}[b]{0.32\textwidth}
            \centering
            \includegraphics[width=\linewidth]{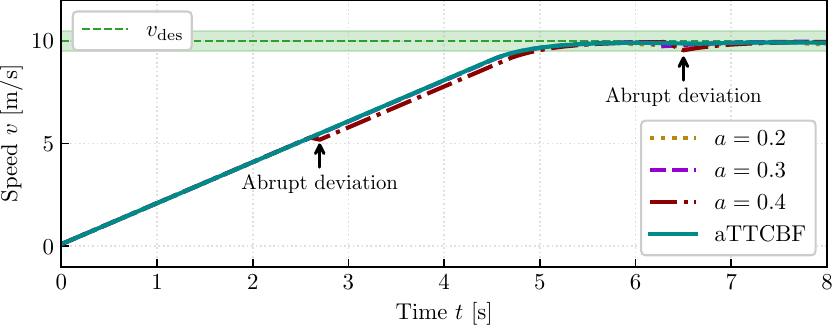}
        \end{minipage}
    }\hfill
    \subfloat[Speed $v$ (exp. $\mathcal{K}$)\label{fig_corridor_attcbf_exp_v}]{
        \begin{minipage}[b]{0.32\textwidth}
            \centering
            \includegraphics[width=\linewidth]{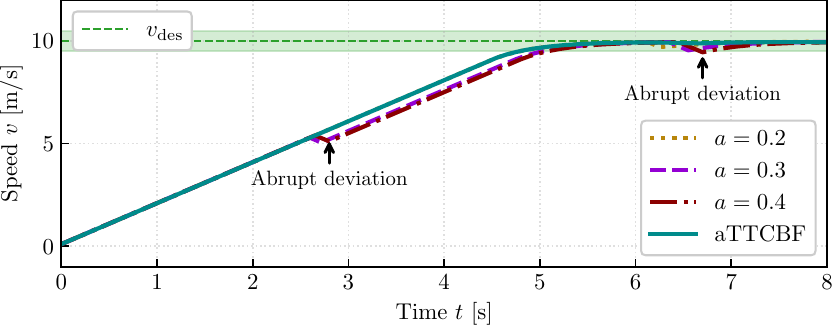}
        \end{minipage}
    }\hfill
    \addtocounter{subfigure}{1} % skip (l)
    \subfloat[Speed $v$ (rational $\mathcal{K}$)\label{fig_corridor_attcbf_rational_v}]{
        \begin{minipage}[b]{0.32\textwidth}
            \centering
            \includegraphics[width=\linewidth]{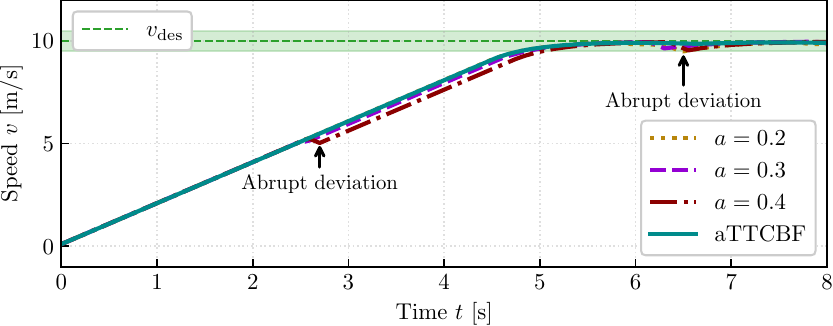}
        \end{minipage}
    }

    \vspace{-3mm}

    % ---------- Row 5: u-v Figures ----------
    \subfloat[Control effort and speed (linear $\mathcal{K}$)\label{fig_corridor_attcbf_linear_u_v}]{
        \begin{minipage}[b]{0.32\textwidth}
            \centering
            \includegraphics[width=\linewidth]{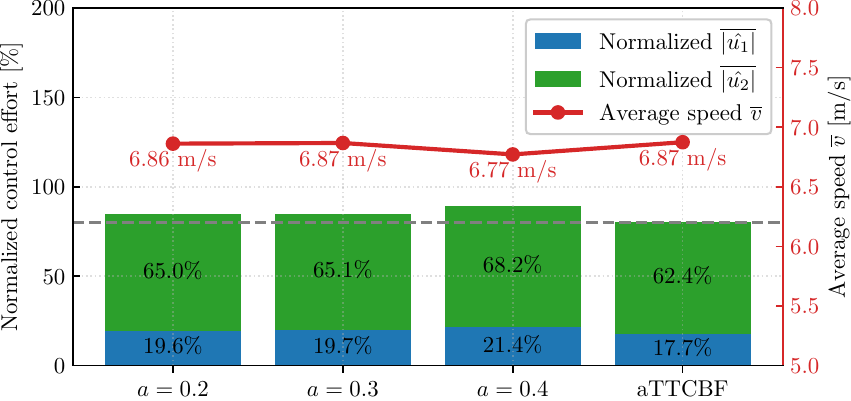}
        \end{minipage}
    }\hfill
    \subfloat[Control effort and speed (exp. $\mathcal{K}$)\label{fig_corridor_attcbf_exp_u_v}]{
        \begin{minipage}[b]{0.32\textwidth}
            \centering
            \includegraphics[width=\linewidth]{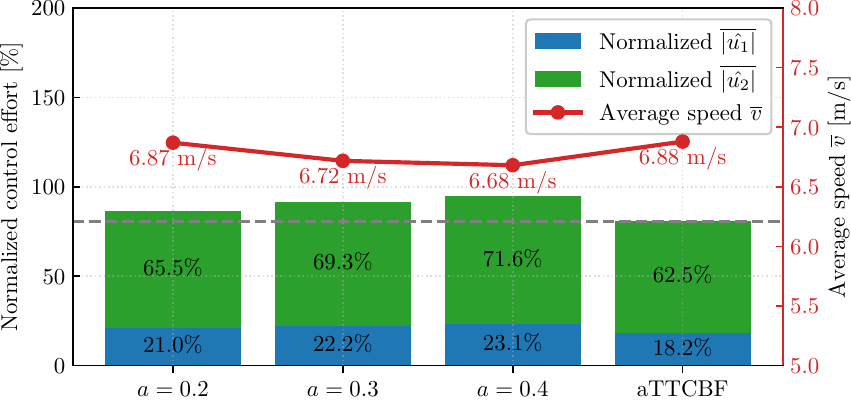}
        \end{minipage}
    }\hfill
    \subfloat[Control effort and speed (rational $\mathcal{K}$)\label{fig_corridor_attcbf_rational_u_v}]{
        \begin{minipage}[b]{0.32\textwidth}
            \centering
            \includegraphics[width=\linewidth]{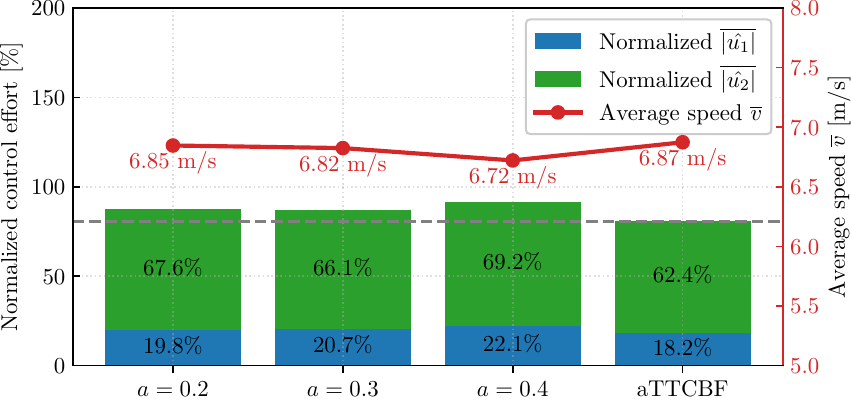}
        \end{minipage}
    }
    \vspace{-2mm}
    \caption{
    Corridor navigation with our TTCBF and aTTCBF with a linear (1st column), an exponential (2nd column), and a rational (3rd column) class $\mathcal{K}$ functions.
    The 1st row: robot trajectories; 
    2nd row: steering rate $u_1$.
    3rd row: acceleration $u_2$.
    4th row: speed $v$.
    5th row: normalized control effort $\overline{|\hat u|}$ and average speed $\overline{v}$.
    }
    \label{fig_corridor_attcbf_main}
\end{figure*}

\begin{figure}[htbp!]
    \setcounter{subfigure}{0}
    \centering

    \subfloat[Robot trajectories.\label{fig_corridor_acbfs_trajectories}]{
        \begin{minipage}[t]{1.0\linewidth}
            \centering
            \includegraphics[width=\linewidth]{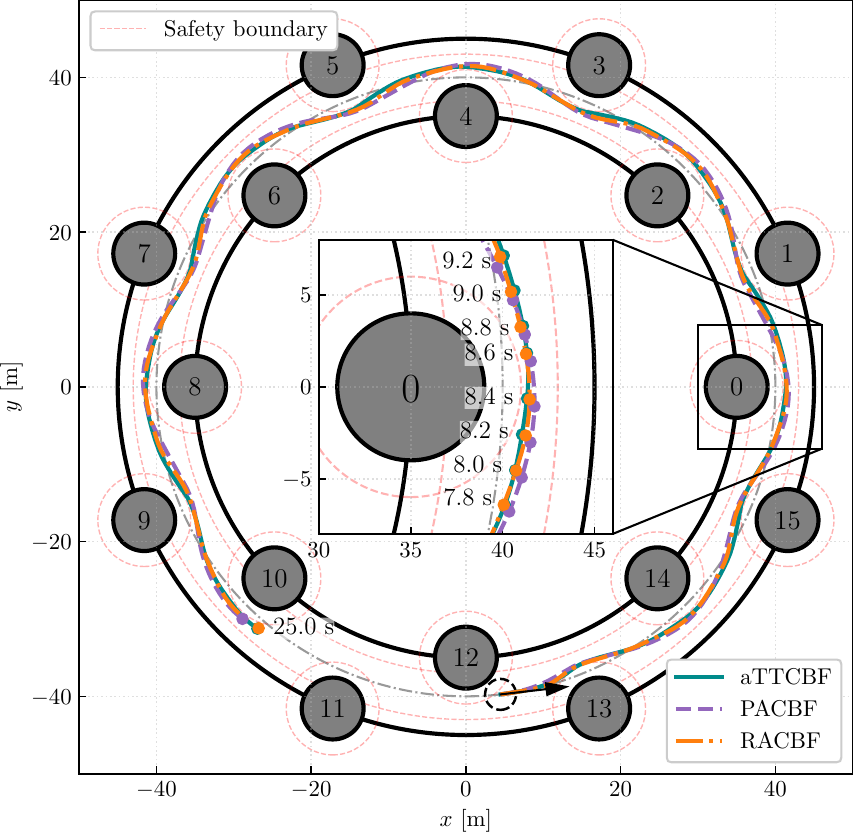}
        \end{minipage}
    }
    \vspace{-0.5pt}
    \subfloat[Steering rate $u_1$.\label{fig_corridor_acbfs_u1}]{
        \begin{minipage}[t]{1.0\linewidth}
            \centering
            \includegraphics[width=\linewidth]{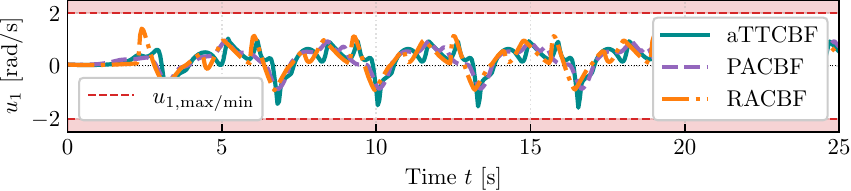}
        \end{minipage}
    }
    \vspace{-0.5pt}
    \subfloat[Acceleration $u_2$.\label{fig_corridor_acbfs_u2}]{
        \begin{minipage}[t]{1.0\linewidth}
            \centering
            \includegraphics[width=\linewidth]{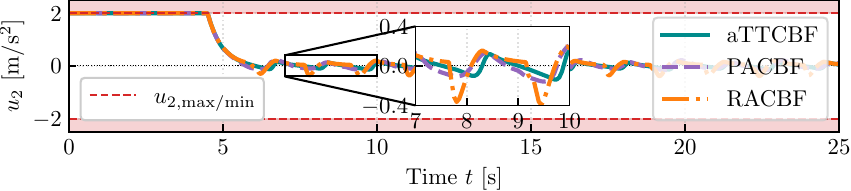}
        \end{minipage}
    }
    \vspace{-0.5pt}
    \subfloat[Speed $v$.\label{fig_corridor_acbfs_v}]{
        \begin{minipage}[t]{1.0\linewidth}
            \centering
            \includegraphics[width=\linewidth]{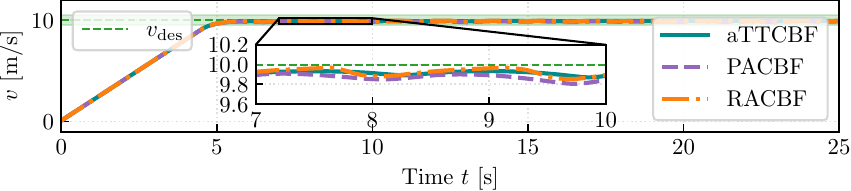}
        \end{minipage}
    }
    \vspace{-0.5pt}
    \subfloat[Path-tracking error $e_{\mathrm{path}}$.\label{fig_corridor_acbfs_path_tracking_error}]{
        \begin{minipage}[t]{1.0\linewidth}
            \centering
            \includegraphics[width=\linewidth]{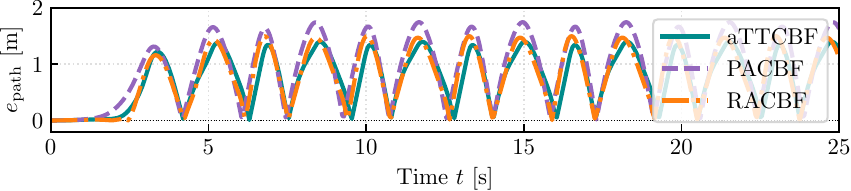}
        \end{minipage}
    }
    \vspace{-2mm}
    \caption{Comparing aTTCBF (our), PACBF, and RACBF: (a) trajectories, (b-c) control inputs, (d-e) control performance.}
    \label{fig_corridor_acbfs_part1}
\end{figure}

\begin{figure}[htbp!]
    \setcounter{subfigure}{0}
    \centering
    \textit{(Continued from Fig.~\ref{fig_corridor_acbfs_part1})}

    \subfloat[Adaptive parameters $\eta(t)$ of our aTTCBF.\label{fig_corridor_acbfs_adaption_attcbf}]{
        \begin{minipage}[t]{\linewidth}
            \centering
            \includegraphics[width=\linewidth]{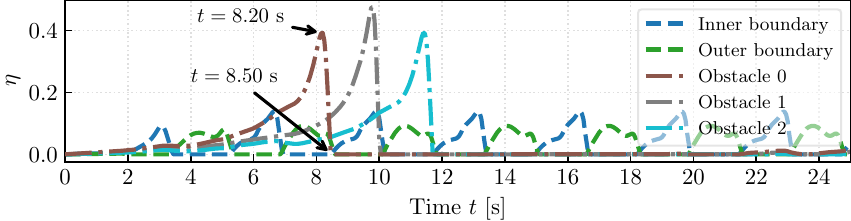}
        \end{minipage}
    }
    \vspace{-0.5pt}
    \subfloat[Adaptive parameters $p_1(t)$ of the PACBF.\label{fig_corridor_acbfs_adaption_pacbf_p1}]{
        \begin{minipage}[t]{\linewidth}
            \centering
            \includegraphics[width=\linewidth]{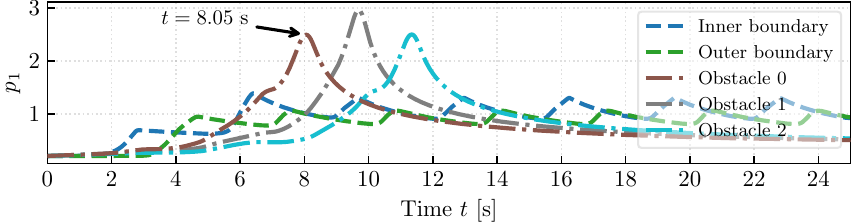}
        \end{minipage}
    }
    \vspace{-0.5pt}
    \subfloat[Adaptive parameters $p_2(t)$ of the PACBF.\label{fig_corridor_acbfs_adaption_pacbf_p2}]{
        \begin{minipage}[t]{\linewidth}
            \centering
            \includegraphics[width=\linewidth]{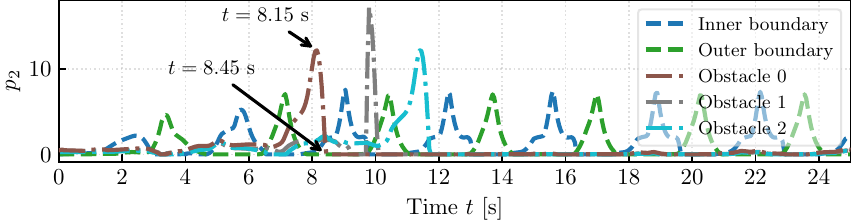}
        \end{minipage}
    }
    \vspace{-0.5pt}
    \subfloat[Adaptive parameters $h_a(t)$ of the RACBF.\label{fig_corridor_acbfs_adaption_racbf}]{
        \begin{minipage}[t]{\linewidth}
            \centering
            \includegraphics[width=\linewidth]{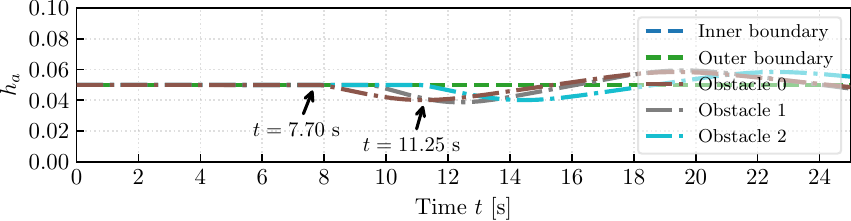}
        \end{minipage}
    }
    \vspace{-2mm}
    \caption{Comparing aTTCBF (our), PACBF, and RACBF: (a-d) adaptive parameters of five selected safety constraints (two corridor boundaries and three obstacles).}
    \label{fig_corridor_acbfs_part2}
\end{figure}

\subsection{Corridor Navigation}\label{sec:eva-corridor}
In Section~\ref{sec:compare-ttcbf-attcbf}, we evaluate how \ac{TTCBF} accommodates different class $\mathcal{K}$ functions in a corridor-navigation scenario and demonstrate the adaptability of \ac{aTTCBF} by comparing it with \ac{TTCBF}. In the same scenario, we further benchmark our \ac{aTTCBF} against two adaptive variants proposed in \cite{xiao2022adaptive}, namely \ac{PACBF} and \ac{RACBF}.

The task is to control a robot to follow the circular corridor centerline (radius \SI{40}{\meter}) counterclockwise at the desired speed $v_{\mathrm{des}}=\SI{10}{\meter\per\second}$, while avoiding collisions with the inner and outer corridor boundaries and with a set of $n_{\mathrm{obs}}=16$ circular obstacles placed along both boundaries, as shown in Fig.~\ref{fig_corridor_acbfs_trajectories}. The robot is modeled with the nonlinear unicycle model
\begin{equation}
    \dot{\bm{x}} =
    \begin{bmatrix}
        \dot{x} \\
        \dot{y} \\
        \dot{\theta} \\
        \dot{v}
    \end{bmatrix}
    =
    \begin{bmatrix}
        v \cos\theta \\
        v \sin\theta \\
        u_1 \\
        u_2
    \end{bmatrix},
    \quad
    \bm{x} =
    \begin{bmatrix}
        x \\ y \\ \theta \\ v
    \end{bmatrix},
    \quad
    \bm{u} =
    \begin{bmatrix}
        u_1 \\ u_2
    \end{bmatrix},
\end{equation}
where $(x,y)$ is the position, $\theta$ the heading, $v$ the longitudinal speed, $u_1$ the steering rate, and $u_2$ the longitudinal acceleration. We use the sampling period $\Delta t=\SI{0.05}{\second}$.

We define the safe sets using signed distances between the robot and the corridor boundaries and between the robot and the obstacles. The resulting \acp{CBF} have relative degree two with respect to the system dynamics. Specifically, for the inner boundary, we use
\[
h=(x-x_{\mathrm{c}})^2+(y-y_{\mathrm{c}})^2-(r_{\mathrm{inn}}+r_{\mathrm{rob}})^2,
\]
and for the outer boundary, we use
\[
h=(r_{\mathrm{out}}-r_{\mathrm{rob}})^2-(x-x_{\mathrm{c}})^2-(y-y_{\mathrm{c}})^2,
\]
where $(x_{\mathrm{c}},y_{\mathrm{c}})=(\SI{0}{\meter},\SI{0}{\meter})$ is the corridor center and
$r_{\mathrm{inn}}=\SI{35}{\meter}$, $r_{\mathrm{out}}=\SI{45}{\meter}$, and $r_{\mathrm{rob}}=\SI{2}{\meter}$
are the radii of the inner boundary, outer boundary, and robot, respectively. For each obstacle $i\in\{0,\ldots,n_{\mathrm{obs}}-1\}$, we define
\[
h=(x-x_i)^2+(y-y_i)^2-(r_{i,\mathrm{obs}}+r_{\mathrm{rob}})^2,
\]
where $(x_i,y_i)$ is the obstacle center and $r_{i,\mathrm{obs}}=\SI{4}{\meter}$ is the obstacle radius. The red dashed circles in Fig.~\ref{fig_corridor_acbfs_trajectories} indicate the zero-level sets of these \acp{CBF}, obtained by inflating the corridor boundaries and obstacles by the robot radius. Any intersection between the robot trajectory and these circles indicates a violation of the safety constraints.

We use two \acp{CLF} to track the corridor centerline (gray dash-dotted lines in Fig.~\ref{fig_corridor_acbfs_trajectories}).
At each time step, we sample a target point on the centerline ahead of the robot by \SI{5.0}{\degree} and define a desired heading $\theta_{\mathrm{des}}$ pointing from the robot toward this point. We set
$V_{\theta}=e_{\theta}^2$ with $e_{\theta}=\theta-\theta_{\mathrm{des}}$ for heading tracking, and
$V_v=e_v^2$ with $e_v=v-v_{\mathrm{des}}$ for speed tracking. The corresponding \ac{CLF} constraints are
$\dot V_v+\lambda_{\mathrm{L},v}V_v\le \zeta_v$ and $\dot V_{\theta}+\lambda_{\mathrm{L},\theta}V_{\theta}\le \zeta_{\theta}$,
where $\dot V_{\theta}=2(\theta-\theta_{\mathrm{des}})u_1$, $\dot V_v=2(v-v_{\mathrm{des}})u_2$,
$\lambda_{\mathrm{L},v}=\lambda_{\mathrm{L},\theta}=4.0$, and $\zeta_{\theta}\ge 0$, $\zeta_v\ge 0$ are \ac{CLF} relaxation variables.
The cost function follows \eqref{eq:qp-cost}. We use a proportional nominal controller
$\bm u_{\mathrm{nom}}=[K_{\theta}e_{\theta},K_v e_v]^\top$ with $K_{\theta}=1.0$ and $K_v=1.0$.
The weights for penalizing the deviation from $\bm u_{\mathrm{nom}}$, the relaxation variables, and the \ac{aTTCBF} adaptive gains are $1$, $100$, and $500$, respectively.

\subsubsection{Comparing aTTCBF with TTCBF}\label{sec:compare-ttcbf-attcbf}
We compare \ac{aTTCBF} with the non-adaptive \ac{TTCBF} under different class $\mathcal{K}$ functions.
For \ac{TTCBF}, the safety constraint is \eqref{eq:thm-ttcbf-cond}, while for \ac{aTTCBF} it is \eqref{eq:thm-attcbf-cond}.
We consider three class $\mathcal{K}$ functions:
1) linear: $\alpha(h)=ah$ for \ac{TTCBF} and $\alpha(h,t)=\eta(t)h$ for \ac{aTTCBF};
2) exponential: $\alpha(h)=ah^{1.1}$ for \ac{TTCBF} and $\alpha(h,t)=\eta(t)h^{1.1}$ for \ac{aTTCBF}; and
3) rational: $\alpha(h)=a\frac{h^2}{1+h}$ for \ac{TTCBF} and $\alpha(h,t)=\eta(t)\frac{h^2}{1+h}$ for \ac{aTTCBF}.
For each type, we test $a\in\{0.2,0.3,0.4\}$, and \ac{aTTCBF} adapts $\eta(t)\in[0,1]$ online. Each simulation runs for \SI{8.0}{\second}.

\textbf{Safety:}
As shown in Figs.~\ref{fig_corridor_attcbf_linear_trajectories}--\ref{fig_corridor_attcbf_rational_trajectories}, both \ac{TTCBF} and \ac{aTTCBF} avoid collisions for all tested parameters, as the trajectories do not intersect the safety boundaries shown by red dashed circles.

\textbf{Speed-tracking performance:}
Across all tested settings, \ac{aTTCBF} yields improved speed tracking relative to \ac{TTCBF}. In
Figs.~\ref{fig_corridor_attcbf_linear_v}--\ref{fig_corridor_attcbf_rational_v}, \ac{aTTCBF} converges smoothly toward $v_{\mathrm{des}}$, whereas \ac{TTCBF} occasionally exhibits abrupt deviations after reaching the desired speed. We also compare the average speed, and an average speed closer to the desired speed $v_{\mathrm{des}}=\SI{10.0}{\meter\per\second}$ indicates better speed tracking. The average-speed curves (red) in
Figs.~\ref{fig_corridor_attcbf_linear_u_v}--\ref{fig_corridor_attcbf_rational_u_v} confirm that \ac{aTTCBF} achieves higher average speed in all cases.

\textbf{Control effort:}
Figs.~\ref{fig_corridor_attcbf_linear_u1}--\ref{fig_corridor_attcbf_rational_u2} show that the control inputs under \ac{aTTCBF} are consistently smoother than under \ac{TTCBF}. For the linear class $\mathcal{K}$ function, Fig.~\ref{fig_corridor_attcbf_linear_u1} shows that the steering rate $u_1$ under \ac{TTCBF} saturates at the upper bound twice, whereas \ac{aTTCBF} avoids these saturations. Similarly, Fig.~\ref{fig_corridor_attcbf_linear_u2} shows that the acceleration $u_2$ under \ac{TTCBF} reaches its lower bound once, while \ac{aTTCBF} does not. Similar behavior is observed for the exponential class $\mathcal{K}$ function in
Figs.~\ref{fig_corridor_attcbf_exp_u1}--\ref{fig_corridor_attcbf_exp_u2} and for the rational class $\mathcal{K}$ function in
Figs.~\ref{fig_corridor_attcbf_rational_u1}--\ref{fig_corridor_attcbf_rational_u2}.
We quantify control effort using the average magnitudes of $u_1$ and $u_2$, normalized by the limits
$u_{1,\max}=\SI{2.0}{\radian\per\second}$ and $u_{2,\max}=\SI{2.0}{\meter\per\second\squared}$, denoted by
$\overline{|\hat{u}_1|}$ and $\overline{|\hat{u}_2|}$. The gray dashed horizontal lines in
Figs.~\ref{fig_corridor_attcbf_linear_u_v}--\ref{fig_corridor_attcbf_rational_u_v} show that \ac{aTTCBF} requires less total control effort across all tested parameters. For example, with the linear class $\mathcal{K}$ function, \ac{aTTCBF} yields $\overline{|\hat{u}_1|}=\SI{17.7}{\percent}$ and $\overline{|\hat{u}_2|}=\SI{62.4}{\percent}$, which are lower than the corresponding values for \ac{TTCBF} across all tested $a$. Similar trends hold for the exponential and rational cases.

In summary, compared with \ac{TTCBF}, \ac{aTTCBF} yields improved speed tracking and reduced control effort across all tested class $\mathcal{K}$ functions. This improvement comes with a modest computational overhead: the average \ac{QP} solving time increases from \SI{0.75}{\milli\second} per step for \ac{TTCBF} to \SI{0.87}{\milli\second} per step for \ac{aTTCBF}, corresponding to a \SI{16.0}{\percent} increase. In Section~\ref{sec:compare-adaptive-cbf}, we show that this overhead is smaller than that of other adaptive variants, namely \ac{PACBF} and \ac{RACBF}.

\subsubsection{Comparing aTTCBF with PACBF and RACBF}\label{sec:compare-adaptive-cbf}
We compare our \ac{aTTCBF} with \ac{PACBF} and \ac{RACBF} from \cite{xiao2022adaptive}. All methods use well-tuned parameters, and each simulation runs for \SI{25.0}{\second}.

\textbf{Safety and control effort:}
Fig.~\ref{fig_corridor_acbfs_trajectories} shows that all three methods avoid collisions. The control inputs are shown in
Figs.~\ref{fig_corridor_acbfs_u1} and \ref{fig_corridor_acbfs_u2}. Overall, the three methods exhibit similar trends. Our \ac{aTTCBF} uses slightly larger negative values of $u_1$ (cyan curve in Fig.~\ref{fig_corridor_acbfs_u1}), while \ac{RACBF} requires noticeably larger negative values of $u_2$ (orange curve in Fig.~\ref{fig_corridor_acbfs_u2}). In terms of control effort, \ac{aTTCBF} yields
$\overline{|u_1|}=\SI{0.45}{\radian\per\second}$ and $\overline{|u_2|}=\SI{0.45}{\meter\per\second\squared}$,
\ac{PACBF} yields
$\overline{|u_1|}=\SI{0.43}{\radian\per\second}$ and $\overline{|u_2|}=\SI{0.47}{\meter\per\second\squared}$,
and \ac{RACBF} yields
$\overline{|u_1|}=\SI{0.44}{\radian\per\second}$ and $\overline{|u_2|}=\SI{0.49}{\meter\per\second\squared}$.
Normalizing by $u_{1,\max}=\SI{2.0}{\radian\per\second}$ and $u_{2,\max}=\SI{2.0}{\meter\per\second\squared}$ and summing the two components, \ac{aTTCBF} and \ac{PACBF} each use \SI{45.0}{\percent} of the control capacity, while \ac{RACBF} uses \SI{46.5}{\percent}.

\textbf{Speed- and path-tracking performance:}
Figs.~\ref{fig_corridor_acbfs_v} and \ref{fig_corridor_acbfs_path_tracking_error} show the speed-tracking performance and the centerline path-tracking error, respectively. The speed profiles are similar for all three methods (Fig.~\ref{fig_corridor_acbfs_v}). The average speeds are \SI{8.94}{\meter\per\second} for \ac{aTTCBF} (\SI{89.4}{\percent} of $v_{\mathrm{des}}$), \SI{8.90}{\meter\per\second} for \ac{PACBF} (\SI{89.0}{\percent}), and \SI{8.94}{\meter\per\second} for \ac{RACBF} (\SI{89.4}{\percent}).
Because the obstacles densely constrain the feasible state space, the robot must deviate from the centerline when avoiding obstacles, so perfect centerline tracking is not achievable. We quantify path tracking using the absolute mean centerline deviation $e_{\mathrm{path}}$. As shown in Fig.~\ref{fig_corridor_acbfs_path_tracking_error}, our \ac{aTTCBF} reduces the mean path-tracking error from \SI{0.95}{\meter} (\ac{PACBF}) and \SI{0.79}{\meter} (\ac{RACBF}) to \SI{0.76}{\meter}, which corresponds to reductions of \SI{20.0}{\percent} and \SI{3.8}{\percent}, respectively.

\textbf{Adaptability:} Our \ac{aTTCBF} uses one adaptive parameter $\eta(t)$. The \ac{PACBF} uses two adaptive parameters, denoted by $p_1(t)$ and $p_2(t)$, while the \ac{RACBF} uses one adaptive parameter $h_a(t)$ \cite{xiao2022adaptive}. Figs.~\ref{fig_corridor_acbfs_adaption_attcbf}--\ref{fig_corridor_acbfs_adaption_racbf} show these parameters for the safety constraints associated with the inner and outer corridor boundaries and three representative obstacles. In Fig.~\ref{fig_corridor_acbfs_adaption_attcbf}, $\eta(t)$ for obstacle 0 (brown dash-dotted line) increases as the robot approaches the obstacle, peaks at $t=\SI{8.20}{\second}$, and decreases to zero at $t=\SI{8.50}{\second}$. This decrease occurs because, once the robot passes the closest point (see the zoomed view in Fig.~\ref{fig_corridor_acbfs_trajectories}) at $t=\SI{8.50}{\second}$, the barrier value stops decreasing and a smaller $\eta(t)$ suffices to maintain the feasibility of the \ac{QP} \eqref{eq:qp}. As the robot moves away, the barrier value increases and the optimizer drives $\eta(t)$ to zero due to the quadratic penalty in \eqref{eq:qp-cost}. Similar behavior is observed for obstacles 1 and 2.

Figs.~\ref{fig_corridor_acbfs_adaption_pacbf_p1} and \ref{fig_corridor_acbfs_adaption_pacbf_p2} show that $p_1(t)$ and $p_2(t)$ exhibit similar trends but can peak at different times. For obstacle 0, $p_1(t)$ peaks at \SI{8.05}{\second}, while $p_2(t)$ peaks at \SI{8.15}{\second}. This behavior illustrates a practical drawback of using multiple adaptive parameters: although the parameters adapt online, the corresponding penalty weights in the cost function must still be tuned, which complicates the control design. We report the best performance obtained after extensive tuning.

Fig.~\ref{fig_corridor_acbfs_adaption_racbf} shows the adaptive parameter $h_a(t)$ for \ac{RACBF}. It is initialized at $0.05$ and regulated toward the same desired value. As observed, $h_a(t)$ decreases at \SI{7.70}{\second} to relax the constraint associated with obstacle 0 and increases again at \SI{11.25}{\second} to return toward the desired value. The adaptation in \ac{RACBF} starts earlier than in \ac{aTTCBF} and \ac{PACBF}, and its response is slower.

\textbf{Computational time and control-design parameters:}
The average \ac{QP} solving time for our \ac{aTTCBF} is \SI{0.71}{\milli\second} per step, which is a \SI{54.5}{\percent} reduction compared to \ac{PACBF} (\SI{1.56}{\milli\second}) and a \SI{9.0}{\percent} reduction compared to \ac{RACBF} (\SI{0.78}{\milli\second}). In addition, in this scenario, our \ac{aTTCBF} requires 18 control-design parameters that are related to class $\mathcal{K}$ functions, which is a \SI{75.0}{\percent} reduction relative to \ac{PACBF} (72 parameters) and a \SI{85.7}{\percent} reduction relative to \ac{RACBF} (126 parameters). We refer to Appendix~\ref{app:design-parameter} for detailed computation.

\begin{table}[t]
\setlength{\tabcolsep}{1.5pt} % default 6pt
\centering
\caption{Overview of the comparison of our aTTCBF, the PACBF~\cite{xiao2022adaptive}, and the RACBF~\cite{xiao2022adaptive}.}
\label{tab:adaptive-cbfs-comparison}
\begin{tabular}{llllll}
\toprule
Methods  & \makecell[l]{Control \\ effort} & \makecell[l]{Speed \\ tracking} & \makecell[l]{Path \\ tracking} & \makecell[l]{\ac{QP}-sol. \\ time} & \makecell[l]{\#Design \\ param.} \\
\midrule
Our aTTCBF & \textbf{45.0}\si{\percent}  & \textbf{89.4}\si{\percent} & \textbf{0.76}\si{\meter} & \textbf{0.71}\si{\milli\second} & \textbf{18} \\
PACBF~\cite{xiao2022adaptive} & \textbf{45.0}\si{\percent} & \SI{89.0}{\percent} & \SI{0.95}{\meter} & \SI{1.56}{\milli\second} & 72 \\
RACBF~\cite{xiao2022adaptive} & \SI{46.5}{\percent} & \textbf{89.4}\si{\percent} & \SI{0.79}{\meter} & \SI{0.78}{\milli\second} & 126 \\
\bottomrule
\end{tabular}
\end{table}

Table~\ref{tab:adaptive-cbfs-comparison} summarizes the performance of our \ac{aTTCBF}, the \ac{PACBF}, and the \ac{RACBF}. Overall, our \ac{aTTCBF} achieves the best speed- and path-tracking performance while requiring the lowest computation time and the fewest control-design parameters. Control effort is comparable between \ac{aTTCBF} and \ac{PACBF}, and both are slightly lower than \ac{RACBF}.

\section{Discussions}\label{sec:discussions}
Compared with the \ac{HOCBF} in \cite{xiao2022highorder}, our \ac{TTCBF} in Section~\ref{sec:ttcbf} offers a key advantage: it accommodates safety constraints with high relative degree using only one class $\mathcal{K}$ function. This yields two benefits. First, the number of class $\mathcal{K}$ functions that require tuning is one, independent of the relative degree of the safety constraint. Second, adaptive variants based on our \ac{TTCBF}, such as our \ac{aTTCBF}, are substantially less complex than existing adaptive variants, including \ac{PACBF} and \ac{RACBF} \cite{xiao2022adaptive}.

A limitation of our \ac{TTCBF} is that the Taylor size $\Delta T=r\Delta t$ (see \eqref{eq:taylor-size}) is coupled with both the controller sampling period $\Delta t$ and the relative degree $r$ of the safety constraint. As shown in \eqref{eq:thm-ttcbf-cond}, the $r$-th Taylor term $h^{(r)}$, which is the first term that depends on the control input, is scaled by $(\Delta T)^r/r!$. When $r$ is large and $\Delta T<\SI{1.0}{\second}$, this coefficient can become very small, which may lead to numerical ill-conditioning in the resulting \ac{QP}. We consider two remedies that increase $\Delta T$. The first is to increase the sampling period $\Delta t$. If a larger $\Delta t$ is not feasible, a second approach is to redefine the Taylor size as $\Delta T=r'\Delta t$ with a chosen integer $r'>r$. 
The second approach introduces a ``look-ahead'' effect to time step $k+r'$ by zero-order holding the control input over the additional steps, that is, by setting $\bm u(t_{k+1}),\ldots,\bm u(t_{k+r'-r})$ equal to $\bm u(t_k)$. This look-ahead mitigates the myopic nature of standard \acp{CBF} \cite{xiao2022highorder}. With an enlarged Taylor size, our \ac{TTCBF} resembles a receding-horizon formulation in which the control horizon is one and the input beyond the horizon is held constant, resembling the idea in \ac{MPC} with a control horizon of one. We have not yet established a theoretical analysis for our \ac{TTCBF} under this enlarged Taylor size and will address it in future work.
% For example, in the spring-mass system in Section~\ref{sec:eva-spring-mass} with relative degree six, we set $\Delta T=r'\Delta t=16\cdot \SI{0.01}{\second}=\SI{0.16}{\second}$. In contrast, for moderate relative degrees such as $r=2$ in the corridor-navigation scenario in Section~\ref{sec:eva-corridor}, we directly use $\Delta T=r\Delta t$. 

%===============================================================================
\section{Conclusions}\label{sec:conclusions}
In this work, we proposed \acl{TTCBF} (\ac{TTCBF}), which generalizes the standard discrete-time \ac{CBF} to consider safety constraints with high relative degree. 
Different from the existing \ac{HOCBF} that employs a chain of class $\mathcal{K}$ functions, our \ac{TTCBF} uses only one. 
Further, we propose an adaptive variant for our \ac{TTCBF}---\acl{aTTCBF} (\ac{aTTCBF}). 
Since the number of tuned class $\mathcal{K}$ functions is independent of the relative degree, our \ac{TTCBF} and \ac{aTTCBF} reduce control-design complexity compared with \ac{HOCBF} and its adaptive variants, \ac{PACBF} and \ac{RACBF}. We successfully demonstrated that our \ac{TTCBF} handles a relative-degree-six safety constraint on a spring-mass system. In a corridor-navigation scenario, we demonstrated that our \ac{TTCBF} and \ac{aTTCBF} are able to accommodate different class $\mathcal{K}$ functions (linear, exponential, and rational). Moreover, our \ac{aTTCBF} outperforms \ac{PACBF} and \ac{RACBF} across multiple metrics, requiring lower average \ac{QP} solving time, lower control effort, and substantially fewer control-design parameters, while improving both speed and path tracking (see Table~\ref{tab:adaptive-cbfs-comparison} for statistics).

% \begin{figure}
% \begin{center}
% \includegraphics[height=4cm]{jcaesar.eps}    % The printed column  
% \caption{Gaius Julius Caesar, 100--44 B.C.}  % width is 8.4 cm.
% \label{fig1}                                 % Size the figures 
% \end{center}                                 % accordingly.
% \end{figure}

% OR

%\begin{figure}
%\begin{center}
%\epsfig{file=jcaesar,width=7cm}
%\caption{Gaius Julius Caesar, 100--44 B.C.}
%\label{fig1}
%\end{center}
%\end{figure}

\begin{ack}                               % Place acknowledgements
This research was supported by the Bundesministerium für Digitales und Verkehr (German Federal Ministry for Digital and Transport) within the project ``Harmonizing Mobility'' (grant number 19FS2035A). 
\end{ack}

\bibliographystyle{plain}        % Include this if you use bibtex 
\bibliography{autosam}           % and a bib file to produce the 
                                 % bibliography (preferred). The
                                 % correct style is generated by
                                 % Elsevier at the time of printing.

\appendix
\section{Computing Control-Design Parameters}\label{app:design-parameter}

We explain how we compute the reported numbers of control-design parameters for aTTCBF, PACBF, and RACBF in Table~\ref{tab:adaptive-cbfs-comparison}.
We define the total number of control-design parameters as
\begin{equation} \label{eq:design-parameters}
n_{\mathrm{design}} := n_w + n_{\mathcal{K}},
\end{equation}
where \(n_w\) and $n_{\mathcal{K}}$ count tunable weighting coefficients in the \ac{QP} cost and tunable parameters of class \(\mathcal{K}\) functions that are introduced by the specific CBF method, respectively.
We do not count tunable parameters shared across all methods (for example, weights used to penalize deviation from the same nominal controller).

In corridor navigation, there are \(n_{\mathrm{obs}}=16\) obstacles and two corridor boundaries, hence the number of safety constraints is $n_{\mathrm{safe}} = 2 + n_{\mathrm{obs}} = 18$. Each safety constraint has relative degree \(r=2\). Each class \(\mathcal{K}\) function carries one scalar tunable parameter.

Our \ac{aTTCBF} introduces one adaptive gain per safety constraint, with each having one associated weight in the \ac{QP} cost. No other class-\(\mathcal{K}\) parameters are tuned because the class \(\mathcal{K}\) function is coefficient-free. Therefore, applying \eqref{eq:design-parameters} with $n_w = n_{\mathrm{safe}}$ and $n_{\mathcal{K}} = 0$ yields
\begin{equation}\label{eq:count_attcbf}
n_{\mathrm{design}} = 18 + 0 = 18.
\end{equation}

\ac{PACBF} introduces \((r-1)\) auxiliary penalty dynamics per safety constraint. Each such auxiliary component contributes
(i) one \ac{QP} weight (for its associated virtual input and/or relaxation term in the method-specific cost) and
(ii) one class-\(\mathcal{K}\) parameter through the method-specific stabilization/safety inequalities.
This yields the method-level counts
$n_w = 2 n_{\mathrm{safe}}(r-1)$ and $n_{\mathcal{K}} = 2 n_{\mathrm{safe}}(r-1)$. With \(n_{\mathrm{safe}}=18\) and \(r=2\),
\begin{equation}\label{eq:count_pacbf_numbers}
n_{\mathrm{design}}
= 2\cdot 18(2-1) + 2\cdot 18(2-1)
= 36 + 36
= 72.
\end{equation}

\ac{RACBF} introduces, per safety constraint, (i) two method-specific \ac{QP} weights and (ii) two coupled HOCBF chains (one for \(h\) and one for the adaptive relaxation), plus one additional class-\(\mathcal{K}\) parameter from the method-specific CLF term used to regulate the relaxation. This gives $n_w = 2 n_{\mathrm{safe}}$ and $n_{\mathcal{K}} = n_{\mathrm{safe}}(2r+1)$, yielding
\begin{equation}\label{eq:count_racbf_numbers}
n_{\mathrm{design}}
= 2\cdot 18 + 18(2\cdot 2 + 1)
= 36 + 90
= 126.
\end{equation}

\end{document}